\documentclass[reprint,prb,aps,floatfix,superscriptaddress]{revtex4-2}

\usepackage[normalem]{ulem}
\usepackage{amsmath}
\usepackage{amsfonts}
\usepackage{graphicx}
\usepackage{chemformula}
\usepackage{mathtools}
\usepackage{tabularx}
\usepackage{color}
\usepackage{hyperref}   

\definecolor{orange}{rgb}{1,0.5,0}

\definecolor{dblue}{rgb}{0, 0.172 , 0.647}

\definecolor{dgreen}{rgb}{0,0.352,0}

\definecolor{brown}{rgb}{0.545, 0.27, 0.0745}

\definecolor{purple}{rgb}{0.502, 0, 0.502}

\definecolor{springgreen}{rgb}{0.196, 0.78, 0.537}

\definecolor{gray}{rgb}{0.4, 0.4, 0.4}

\begin{document}


\title{Exact Solution for the Transverse Field Sherrington-Kirkpatrick Spin Glass Model with Continuous-Time Quantum Monte Carlo Method}

\author{Annam\'aria Kiss}
\affiliation{Institute for Solid State Physics and Optics, Wigner Research Centre for Physics, H-1525 Budapest, P.O.B. 49, Hungary}
\author{Gergely Zar\'and}
\affiliation{Department of Theoretical Physics, Budapest University of Technology and Economics, Budapest H-1521, Hungary}
\author{Izabella Lovas}
\affiliation{Kavli Institute for Theoretical Physics, University of California, Santa Barbara, 93106, California, USA}


\begin{abstract}
We construct  the first complete exact numerical solution of a mean field quantum spin glass model, the transverse field Sherrington-Kirkpatrick model, by implementing a continuous-time quantum Monte Carlo method in the presence of full replica symmetry breaking. We extract the full numerically exact phase diagram, displaying a glassy phase with continuous replica symmetry breaking at small transverse fields and low temperatures. A paramagnetic phase emerges once thermal and quantum fluctuations melt the spin glass. We characterize both phases by extracting the order parameter, as well as the static and dynamical local spin susceptibilities. The static susceptibility shows a plateau in the glassy phase, but remains smooth across the phase boundary, while the shape of dynamical susceptibility varies upon crossing the glass transition by reducing quantum fluctuations. We qualitatively compare these results to a.c. susceptibility measurements on dipole-coupled Ising magnets in a transverse magnetic field. Our work provides a general framework for the exact numerical solution of mean field quantum glass models, constituting an important step towards understanding glassiness in realistic systems.

\end{abstract}

\maketitle
 
\section{Introduction}

Understanding the interplay of quantum fluctuations and glassiness in spin systems with frustrated interactions is a challenging unresolved problem in condensed matter physics~\cite{tanaka2017quantum}. These systems, known as quantum spin glasses~\cite{bray1980replica}, have regained attention in recent years due to their relevance in quantum annealing\cite{santoro2002theory,zhu2016best, hauke2020perspectives}, or by offering new possibilities for efficiently solving combinatorial optimization problems such as traveling salesman problem~\cite{hen2011exponential, mohseni2022ising,stadler1992landscape} or the graph partitioning problem~\cite{fu1986application,goldschmidt1990replica}, as well as for their potential use as quantum neural networks~\cite{amit1985spin,dotsenko1994introduction,gardner1987zero,QED_PhysRevX.11.021048}.
Despite being of fundamental importance and having a long history in solid state research, the physics of quantum spin glasses still poses numerous open questions.
In particular, the properties of the glassy phase remain extremely challenging, due to the complexity of the problem. 

Classical spin glasses have been extensively intestigated 
and much progress has been achieved by obtaining through the exact solution of mean field models, such as the famous classical Sherrington-Kirkpatrick (SK) model~\cite{sherrington1975solvable,de1978stability,mezard1987spin,panchenko2013sherrington}. The succesfull strategy of focusing on exactly solvable mean field classical models has the potential to shed light on the effect of quantum fluctuations as well, and pave the way to study dynamical properties of quantum spin glasses. 
However, previous studies that followed this route and relied on various approximations, such as static approximations~\cite{yamamoto1987perturbation,kopec1989instabilities},  Landau theory close to the phase transition~\cite{read_sachdev_ye1995, caltagirone2012critical,young2017stability}, and even Monte Carlo study~\cite{rozenberg_grempel1998} showed that it is difficult to accurately treat the complex interplay between frustrated interactions and quantum fluctuations even at the mean field level.

In this regard, the simplest possible extension that adds quantum mechanical aspects to the celebrated SK model is the transverse-field Sherrington-Kirkpatrick model.  This model is an extension of the classical  SK model  by introducing a transverse magnetic field that causes quantum fluctuations in the system~\cite{ray1989sherrington,ishii1985effect}. 

These quantum fluctuations tend to supress the spin glass transition by competing with the effect of the random Ising interactions, producing rich physics in the glassy phase.

The phase diagram in the temperature - transverse field plane has been studied extensively in the last three decades~\cite{yamamoto1987perturbation, ray1989sherrington, LaiGoldschmidt1990, ButtnerUsadel1990, read_sachdev_ye1995, young2017stability, MukherjeeRajakChakranarti2015, MukherjeeRajakChakranarti2018}.
 Bray and Moore~\cite{bray1980replica} reduced the quantum spin glass problem to a single site problem with a time-dependent self interaction term using replica theory, which opened ways for successful studies of the properties of the spin glass state.
There are studies on the critical behaviour near the $T=0$ quantum critical point~\cite{miller_huse1993}, on the paramagnetic solution considered in the glassy phase as well~\cite{rozenberg_grempel1998}, and
even on the glass phase in some extent in the ground state~\cite{rozenberg_arrachea2001, Andreanov2012}.
However, with our best knowledge, there is no full exact solution of the transverse-field Sherrington-Kirkpatrick model is exist in the full parameter range.

In this work we provide the first full exact numerical solution of the transverse-field Sherrington-Kirkpatrick model,
\begin{equation}
\hat{{\cal H}} = -\sum_{(i,j)} J_{ij} \hat{\sigma}^{z}_{i} \hat{\sigma}^{z}_{j} - \sum_{i=1}^N h_{i} \hat{\sigma}^{z}_{i} - h_{T} \sum_{i=1}^N \hat{\sigma}^{x}_{i}.
\label{eq:hamiltonian}
\end{equation}
Here $\{\sigma_i^z,\sigma_i^x\}_{i=1}^N$ denote the spin Pauli matrices on $N$ sites. 
The first term in $\hat{{\cal H}}$ accounts for all-to-all Ising interactions between the  $N(N-1)/2$ pairs of spins, with the Ising couplings $J_{ij}$ chosen as independent Gaussian random variables with zero mean, $\langle J_{ij} \rangle =0$, and variance $\langle J_{ij}^2 \rangle =J^2/N$, with $J$ setting the typical interaction strength. The second term in Eq.~\eqref{eq:hamiltonian} encodes a random site-dependent magnetic field pointing into the $z$ direction, with independent Gaussian variables $h_{i}$ also having a zero mean, $\langle h_{i} \rangle =0$, and a variance $\langle h_{i}^2 \rangle =h^{2}_{z}$. These two terms define the classical Sherrington-Kirkpatrick model.  The third term adds a transverse field $h_T$, pointing in the $x$ direction, that introduces quantum fluctuations. 

\begin{figure}[t]
\centering
\includegraphics[width=0.9\columnwidth]{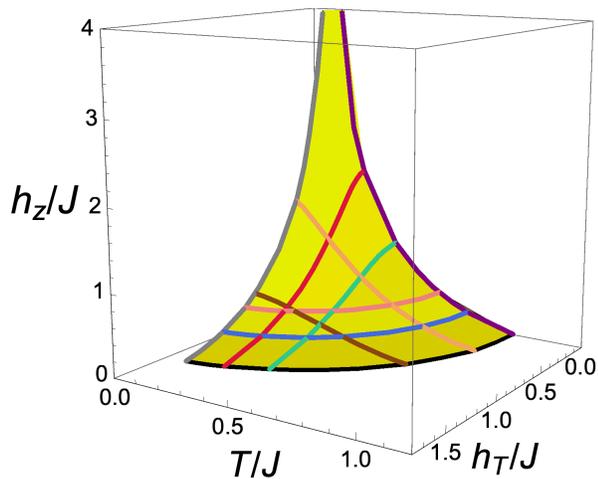}
\caption{Complete phase diagram in terms of  transverse field $h_{\rm T}$, temperature $T$, and onsite disorder $h_{z}$, measured  in units of the interaction strength $J$. The phase boundary separates a quantum spin glass phase with full replica symmetry breaking at low $T$ and $h_T$, while larger thermal and quantum fluctuations induce a paramagnetic phase.
Lines with different colors indicate the cuts shown in Fig.~(\ref{fig-phase-diagrams}).}
\label{fig-3D-phase-diagram}
\end{figure}

We combine the dynamical mean field theory (DMFT)~\cite{georges1996dynamical, kotliar2006electronic} and the continuous-time quantum Monte Carlo (CTQMC) method~\cite{wernerCTQMCPRL2006, wernerCTQMCPRB2006}
to obtain the full exact numerical solution of the  transverse-field Sherrington-Kirkpatrick model ~\eqref{eq:hamiltonian} in the presence of complete replica symmetry breaking. 
Our results are summarized in Fig.~\ref{fig-3D-phase-diagram}.

Combining the two approaches allows us to explore the complete phase diagram, 
and examine the characteristics of the glassy phase, including local static and dynamical spin susceptibilities and the distribution of overlaps between different replicas.  Our mean field results for dynamical susceptibility are compared with experimental measurements performed on the rare-earth compound LiHo$_x$Y$_{1-x}$F$_4$ in a transverse field, which is believed to undergo a quantum spin glass transition~\cite{wu1991classical,Wu_Aeppli1993, Brooke_Aeppli_Sience1999,ghosh2002coherent,torres2008quantum,quilliam2008evidence,quilliam2012exp}.  We find a good qualitative agreement between our results and the experimental data. We also show that our framework can be generalized to obtain the exact solution of mean field electron glass models, therefore, it serves as an important stepping stone towards understanding glassiness in a wider range of physical systems.

The paper is structured in the following way. Section~\ref{sec:DMFT} introduces the DMFT mapping of the lattice model to a local effective action. We first discuss DMFT within the framework of the cavity approach in Sec.~\ref{subsec:cavity}. We then  present a more formal replica method in Sec.~\ref{subsec:replica}, including details on the replica symmetric solution and full replica symmetry breaking. Section~\ref{sec:ctqmc} provides details on the continuous-time quantum Monte Carlo approach employed. 
Section~\ref{sec:numerics} presents our numerical results on the full phase diagram as well as on the properties of the glassy phase including the local and dynamical spin susceptibilities, order parameters, and the distribution of the local effective magnetic field.
Section~\ref{sec:experiment} compares the theoretical findings with experimental data. We present an outlook to electron glasses in Sec.~\ref{sec:electron}, and discuss our results in Sec.~\ref{sec:conclusions}. 
Technical details are relegated to the Appendices.

\section{Dynamical mean-field theory and the replica scheme} \label{sec:DMFT}

In the thermodynamic limit $N\to\infty$, the spin Hamiltonian ~\eqref{eq:hamiltonian} can be mapped exactly to a single site effective action by applying dynamical mean field theory. 
The mean field equations can be derived  either by using the replica method 
or by the cavity approach.
However, to our knowledge the mean-field equations have never been solved before in their full power for the complete phase diagram. 

Before turning to the somewhat technical replica method, we first discuss the more intuitive cavity approach, shedding light on the structure of the mean field equations to be solved.

\subsection{Cavity approach and effective local action}\label{subsec:cavity}

In the cavity approach, we consider the action corresponding to Hamiltonian ~\eqref{eq:hamiltonian},
\begin{equation}
S  = \int_\tau  \sum_{i=1}^N\left(h_{i}\sigma^z_{i\,\tau}+h_T\,\sigma^x_{i\tau}\right)+\sum_{(i,j)}J_{ij}\int_\tau \sigma^z_{i\,\tau}\sigma^z_{j,\tau},\label{eq:action}
\end{equation}
determining the partition function $Z$ through the path integral
\begin{equation}
    Z=\int\mathcal{D}\sigma^z e^{-S}.\label{eq:Z}
\end{equation}
Here we used the shorthand notation $\int_\tau = \int_0^\beta d\tau$, with $\beta$ denoting the inverse temperature, and the path integral $\int\mathcal{D}\sigma^z$ stands for a summation running over all possible spin $z$ trajectories $\{\sigma_{j\tau}^z\}_{j=1...N}^{0\leq\tau\leq\beta}$, with $\sigma_{j\tau}^z=\pm 1$. The transverse field $h_T$ allows spin flip processes connecting the spin $z$ configuration at times $\tau$ and $\tau+\Delta\tau$ through the matrix element
\begin{gather}
    \langle \{\sigma_{j\tau+\Delta\tau}^z\}_{j=1...N}| \,e^{\Delta\tau\, h_T\,\sigma_i^x}\,|\{\sigma_{j\tau}^z\}_{j=1...N}\rangle\approx\phantom{aaa}\nonumber \\
    \quad\quad\left(\delta_{\sigma^z_{i\tau+\Delta\tau},\sigma^z_{i\tau}}+\Delta\tau\, h_T\,\delta_{\sigma^z_{i\tau+\Delta\tau},-\sigma^z_{i\tau}}\right)\prod_{j\neq i}\delta_{\sigma^z_{j\tau+\Delta\tau},\sigma^z_{j\tau}},
\end{gather}
with $\delta$ standing for the Kronecker delta function. We condensed our notation by abbreviating the product of such  matrix elements as $\exp(h_T\int_\tau \sigma_{i\tau}^x)$.
We focus on a single site $i=0$, and divide the action as follows,
\begin{equation}\label{eq:S}
S  = \int_\tau  \left(h_0\sigma^z_{0\,\tau}+h_T\,\sigma^x_{0\tau}\right)+\sum_{j:j\neq 0}J_{0j}\int_\tau \sigma^z_{j\,\tau}\sigma^z_{0\,\tau} + S_{j\ne0}
\;.
\end{equation}
Here the first term describes an isolated spin at site 0, the second term accounts for the coupling between the spin at site 0 and the rest of the system, while $S_{j\ne0}$ collects all terms not involving site 0. Expanding the partition function in terms of the couplings $J_{0j}$, and integrating out the sites $j\neq 0$ leads to a local effective action,
 \begin{gather}
S_0^{\rm eff}  =  \int_\tau  \left(h_0\,\sigma^z_{0\,\tau}+h_T\,\sigma^x_{0\tau}\right)+\int_\tau \sigma^z_{0\,\tau}\sum_{j:j\neq 0}\langle J_{0j}\sigma^z_{j\,\tau} \rangle_{\rm cav}\nonumber\\
-\dfrac{J^2}{2N}\int_\tau\int_{\tau^\prime}\sigma^z_{0\,\tau}\sigma^z_{0\,\tau^\prime}\sum_{j:j\neq 0}\langle\sigma^z_{j\,\tau}\sigma^z_{j\,\tau^\prime} \rangle_{\rm cav},
\end{gather}
with $\langle ...\rangle_{\rm cav}$ denoting cavity expectation values, calculated in the absence of site 0, i.e., with action $S_{j\ne0}$. All higher order terms vanish in the thermodynamic limit $N\to\infty$. The second term in the effective action describes the mean field renormalization of the $z$ magnetic field $h_0$ due to Ising interactions, whereas integrating out the bath also provides time non-local interactions, encoded in the third term of $S_0^{\rm eff}$. By introducing the renormalized $z$ field $\tilde{h}_0$, and the spatial average of the cavity dynamical susceptibilities,
\begin{equation}
    \chi(\tau-\tau^\prime)\equiv\dfrac{1}{N}\sum_{j:j\neq 0}\langle\sigma^z_{j\,\tau}\sigma^z_{j\,\tau^\prime} \rangle_{\rm cav},\label{eq:chi}
\end{equation}
we obtain
\begin{gather}
    S_0^{\rm eff}  = \int_\tau \left(\tilde{h}_0\sigma^{z}_{0\tau}+h_{T}\,\sigma^{x}_{0\tau}\right)
     - \frac{J^2}{2} \int_{\tau, \tau^{\prime}}\chi(\tau-\tau^{\prime}) \sigma^{z}_{0\tau} \sigma^{z}_{0\tau^{\prime}}.
    \label{eq:S_0_eff}
\end{gather}

Dynamical mean field theory therefore results in an ensemble of local actions, parametrized by the random magnetic field $\tilde{h}_0$, such that the spatial average over the original lattice sites is replaced by an average over $\tilde{h}_0$. However, determining the distribution of $\tilde{h}_0$, $P(\tilde{h}_0)$, is challenging, since the Gaussian distribution of the bare magnetic field $h_0$ is renormalized by the Ising interactions. While its shape remains Gaussian in the paramagnetic phase at high temperatures or transverse fields, in  the glassy phase it acquires a more complicated structure. In this case, it is convenient to rely on the replica method, introduced in the next subsection, allowing us to systematically determine the distribution  $P(\tilde{h}_0)$ from arguments reminiscent of a renormalization group procedure. This more formal approach leads to a local effective action with the same structure as the one obtained in the cavity method, but also provides a closed set of equations for $P(\tilde{h}_0)$.

\subsection{Replica method}\label{subsec:replica}

The replica method is the usual technique to study the spin glasses~\cite{mezard1987spin}. It involves 
to replicate the system into multiple replicas (copies), and introducing a set of order parameters 
that describes the correlation between the replicas, in order to evaluate the free energy or 
the correlation functions of the underlying model. 

In the replica approach,  the logarithm of the partition function is rewritten as
\begin{equation}\label{eq:logZ}
\log Z =\lim_{n\rightarrow 0}\dfrac{Z^n-1}{n}\;.
\end{equation}
This formula can be interpreted as introducing  $n\to 0$ copies of the Hamiltonian. The disorder averaged free energy can now be determined by performing the averaging over the Gaussian variables $h_i$ and $J_{ij}$ in $Z^n$, leading to an effective action connecting different replicas. Similarly to the calculation presented in Sec.~\ref{subsec:cavity}, we can also integrate out all sites with the exception of site 0, leading to a local, replicated effective action,
\begin{align}\label{eq:Srep}
    &S_{\rm rep}= \sum_{a=1}^n\left[\int_\tau h_T\,\sigma^x_{a\tau}-\dfrac{J^2}{2}\int_\tau\int_{\tau^\prime}\chi(\tau-\tau^\prime)\sigma^z_{a\,\tau}\sigma^z_{a\,\tau^\prime}\right]\nonumber\\
    &\quad-\sum_{a,b=1}^n\dfrac{h_z^2}{2}\int_\tau\int_{\tau^\prime}\sigma^z_{a\,\tau}\sigma^z_{b\,\tau^\prime}-\dfrac{J^2}{2}\sum_{a\neq b}^n Q_{ab}\int_\tau\int_{\tau^\prime}\sigma^z_{a\,\tau}\sigma^z_{b\,\tau^\prime}.
\end{align}
Here we dropped the label 0 for the site in the local action, and introduced the replica indices $a,b\leq n$. The parameters of the replicated action $S_{\rm rep}$ are determined through the self-consistency conditions,
\begin{align}\label{eq:selfconsistent}
    &\chi(\tau-\tau^\prime)= \langle\sigma^z_{a\,\tau}\sigma^z_{a\,\tau^\prime}\rangle_{S_{\rm rep}},\nonumber\\
    &Q_{a\neq b}=\langle\sigma^z_{a\,\tau}\sigma^z_{b\,\tau^\prime}\rangle_{S_{\rm rep}},
\end{align}
with the expectation values calculated with respect to $S_{\rm rep}$. We note that the spin correlations between different replicas, encoded in $Q_{a\neq b}$, remain static, in contrast to the replica diagonal correlator $\chi(\tau)$. This property reflects the fact that replica off-diagonal correlations are generated by the static disorder, due to  the same disorder configuration being shared by all replicas~\footnote{This follows from noting that replicas only become coupled upon performing the disorder average. Therefore, replica off-diagonal correlators in a fixed disorder configuration factorize, and $\langle\sigma^z_{a\,\tau}\sigma^z_{b\,\tau^\prime}\rangle_{S_{\rm rep}}$ can be expressed as the disorder average of $\langle\sigma^z_{a\,\tau}\rangle_{\{h_i,J_{ij}\}}\langle\sigma^z_{b\,\tau^\prime}\rangle_{\{h_i,J_{ij}\}}$, with both expectation values calculated with the unaveraged replicated action for a fixed parameter set $\{h_i,J_{ij}\}$. In a static disorder and in equilibrium, both factors are time independent, resulting in a static $Q_{ab}$.}.  

The coupling between different replicas in $S_{\rm rep}$ can give rise to spontaneous replica symmetry breaking, signaling an ergodicity breaking glass transition. Before discussing this most general, replica symmetry breaking solution, we first 
address  the paramagnetic, replica symmetric phase in the following subsection.

\subsubsection{Replica symmetric solution}\label{subsec:rs}

In the replica symmetric solution it is assumed that the permutation symmetry between replicas remains unbroken, i.e., $Q_{a\neq b}\equiv Q_{RS}$, with $Q_{RS}$ encoding the overlap between an arbitrary pair of (different) replicas. As we will discuss later in more detail, this assumption is valid in the paramagnetic phase, at high temperatures or transverse fields.

This replica symmetric Ansatz allows us to decouple the different replicas in Eq.~\eqref{eq:Srep}, at the expense of introducing a Hubbard-Stratonovich field $y$. The resulting replica diagonal action has the same structure as $S_0^{\rm eff}$ obtained through the cavity approach, and is given by
\begin{equation}\label{eq:S_y}
S(y) = \int_\tau  \left(y\,\sigma^z_{\tau}+h_T\,\sigma^x_{\tau}\right) - \frac{J^2}{2} \int_{\tau} \int_{\tau^{\prime}} \tilde{\chi}(\tau-\tau^{\prime}) \sigma^{z}_{\tau} \sigma^{z}_{\tau^{\prime}}.
\end{equation}
The Hubbard-Stratonovich field $y$ appearing in this action can be interpreted as a renormalized $z$ magnetic field, incorporating the bare disordered field, and the effect of Ising interactions on the mean field level. In the presence of replica symmetry, its distribution $P_{RS}(y)$ retains the Gaussian form of bare disorder, with a variance renormalized by the interactions,
\begin{equation}\label{eq:PRS}
    P_{RS}(y)=\dfrac{1}{\sqrt{2\pi\left(h_z^2+J^2Q_{RS}\right)}}\exp\left(-\dfrac{y^2}{2\left(h_z^2+J^2Q_{RS}\right)}\right),
\end{equation}
with replica offdiagonal overlap $Q_{RS}$ determined self-consistently as
\begin{equation}\label{eq:QRS}
    Q_{RS}=\int dy\, P_{RS}(y)\,\langle\sigma^z\rangle_{S(y)}^2.
\end{equation}
Finally, the time non-local coupling $\tilde{\chi}(\tau-\tau^{\prime})$ in Eq.~\eqref{eq:S} is expressed as the field averaged connected spin correlator,
\begin{equation}\label{eq:chiRS}
    \tilde{\chi}(\tau)=\int dy\, P_{RS}(y)\,\tilde{\chi}_y(\tau)
\end{equation}
with 
\begin{equation}
\tilde{\chi}_y(\tau-\tau^{\prime})=\langle\sigma^z_{\tau}\sigma^z_{\tau^\prime}\rangle_{S(y)}-\langle\sigma^z\rangle_{S(y)}^2.\label{eq:tilde_chi_y}
\end{equation}
 We note that $\tilde{\chi}(\tau)$ introduced in  Eq.~\eqref{eq:S_y} is given by 
 \begin{equation}
    \tilde{\chi}(\tau)=\chi(\tau)-Q_{RS}\label{eq:tilde_chi}
 \end{equation}
in terms of the replica diagonal correlator defined in Eq.~\eqref{eq:selfconsistent}.

The self-consistency problem ~\eqref{eq:S_y}-\eqref{eq:chiRS} can be solved iteratively, by applying the CTQMC method. First, we initialize the $Q_{RS}^{[0]}$ and $\tilde\chi^{[0]}(\tau)$. Then, at each step $i$ of the iteration, we calculate $\langle\sigma^z\rangle_{S(y)}$, as well as the correlator $\tilde{\chi}_y(\tau)$, on a fine enough grid in $y$ with the CTQMC approach, using $\tilde\chi^{[i]}(\tau)$ as the parameter of the action $S(y)$. We then set $P_{RS}(y)$ by substituting $Q_{RS}^{[i]}$ into Eq.~\eqref{eq:PRS}, and update the parameters for the next iteration step, by calculating $\tilde\chi^{[i+1]}(\tau)$ from Eq.~\eqref{eq:chiRS} and $Q_{RS}^{[i+1]}$ from Eq.~\eqref{eq:QRS}. This procedure is repeated until convergence.

We leave the details of solving the action ~\eqref{eq:S_y} with CTQMC to Sec.~\ref{sec:ctqmc}. Instead, we first discuss the most general solution of Eqs.~\eqref{eq:Srep} and ~\eqref{eq:selfconsistent}, capturing full replica symmetry breaking in the spin glass phase.

\subsubsection{Full replica symmetry breaking}\label{subsec:rsb}

\begin{figure*}[t]
\centering
\includegraphics[width=0.6\columnwidth]{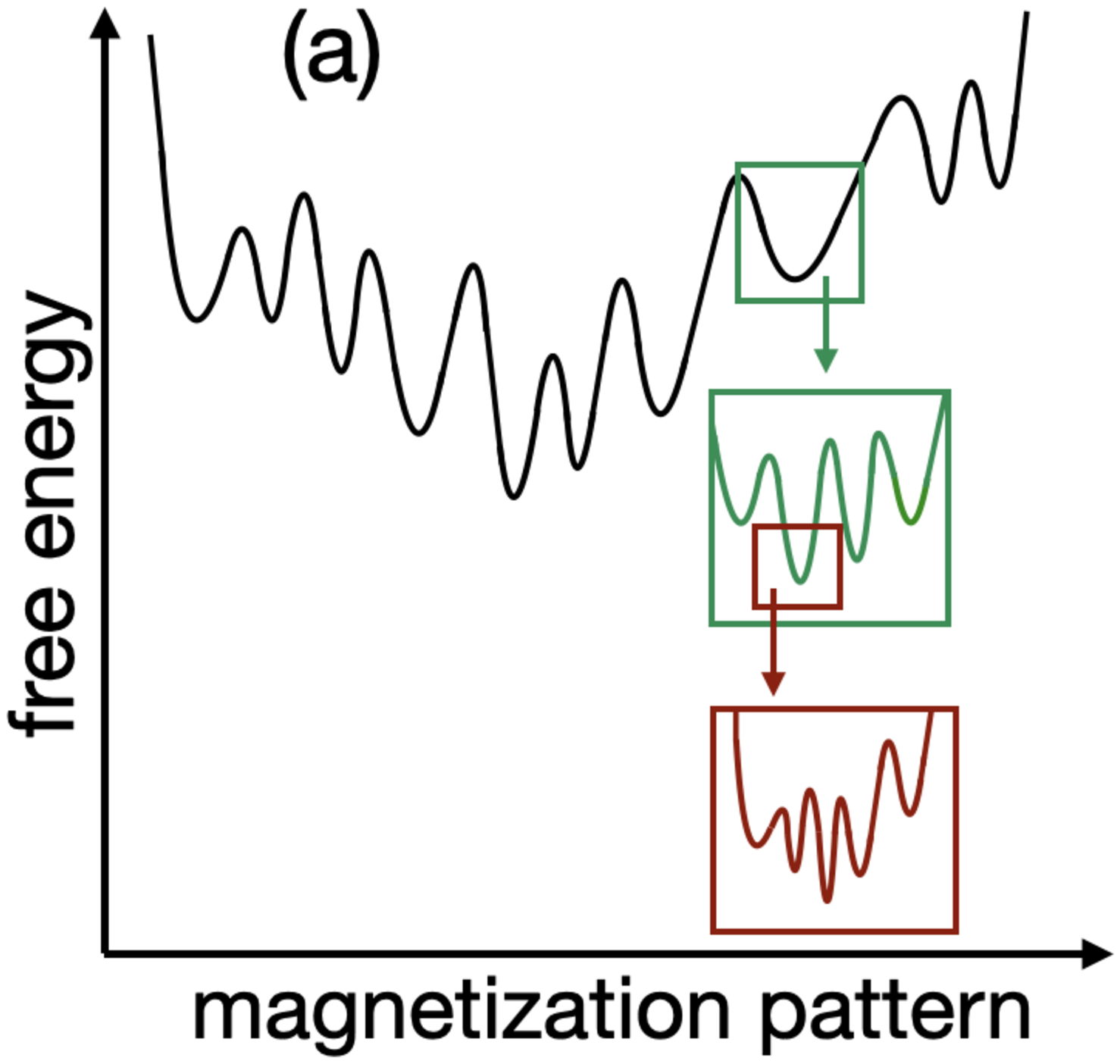}\hspace*{0.4cm}
\includegraphics[width=0.6\columnwidth]{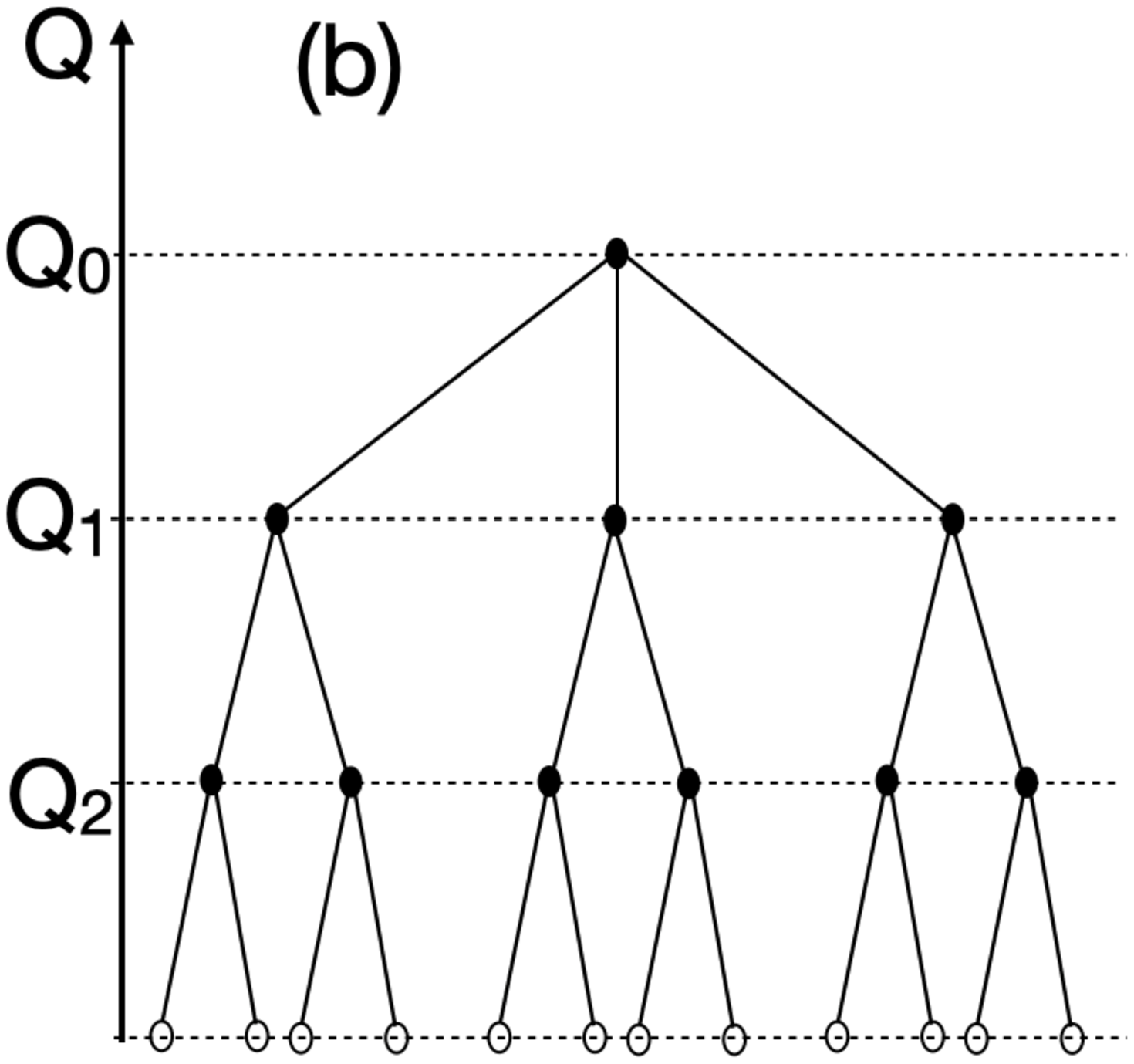}\hspace*{0.4cm}
\includegraphics[width=0.6\columnwidth]{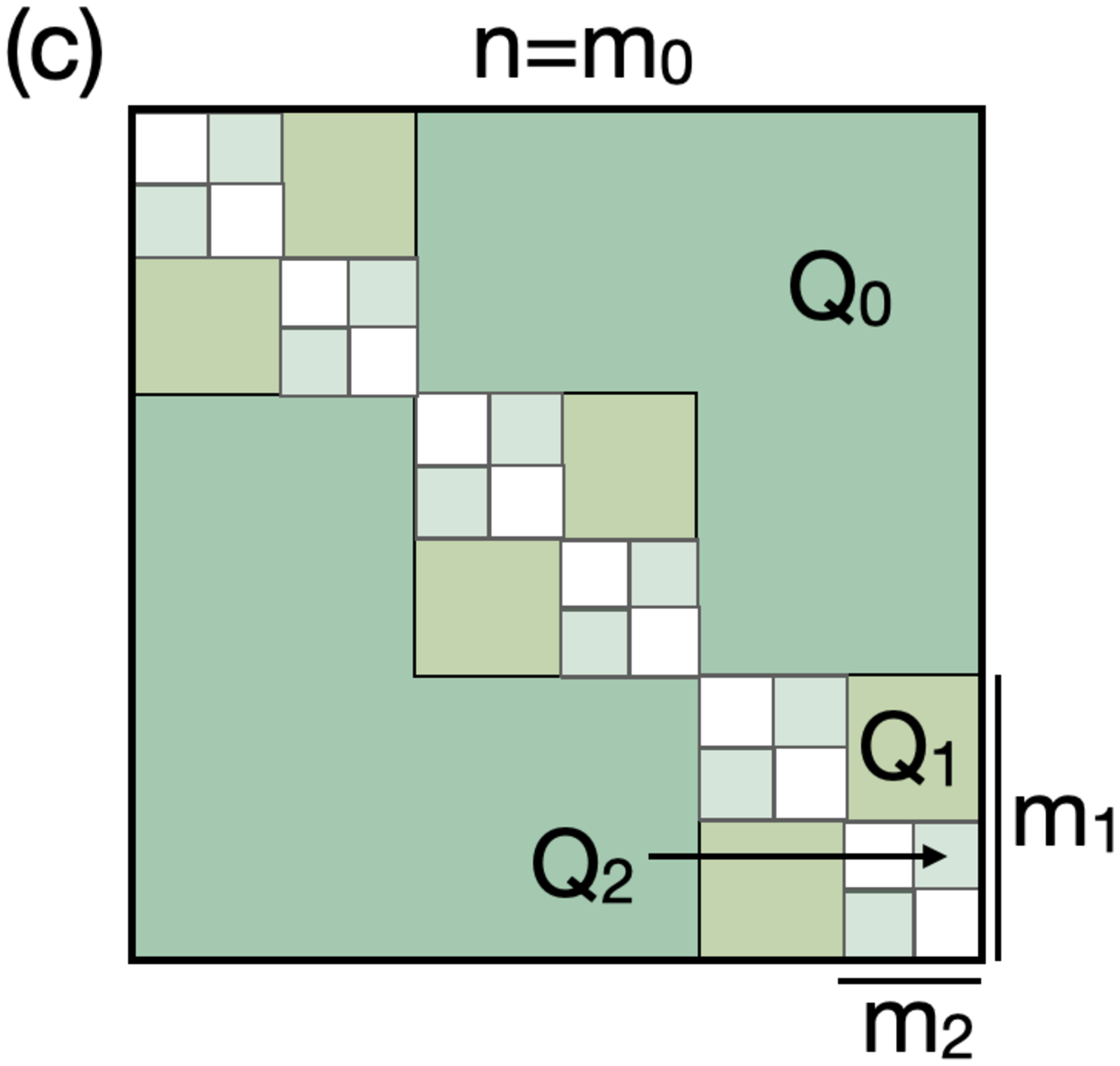}
\caption{
(a) Schematic plot of a rough, glassy free energy landscape. Free energy displays a plethora of metastable minima, corresponding to different magnetization patterns. A Parisi spin glass state is characterized by a hierarchy of free energy valleys, such that larger valleys with states sharing the same coarse grained magnetization keep fracturing into a set of smaller valleys (see zoom-ins), differing in magnetization on shorter scales. (b)-(c) Manifestation of the rough free energy landscape in abstract replica space. (b) Hierarchical tree of replicas, representing the replica overlaps $Q_0\leq Q_1\leq Q_2\leq...$. Leafs at the bottom denote the $n$ replicas. Overlap between two replicas is $Q_{m}$ if the smallest branch shared by them starts at level $m$. (c) Parisi overlap matrix $Q_{ab}$, with a nested block diagonal structure reflecting the hierarchical tree. For illustration we show a  two-step replica symmetry breaking. 
}
\label{fig_free_energy_landscape}
\end{figure*}

To enter the spin glass phase, we have to allow for full replica symmetry breaking in the solution of Eqs.~\eqref{eq:Srep} and ~\eqref{eq:selfconsistent}, i.e., for a non-trivial replica index dependence in the overlap matrix $Q_{ab}$. This replica symmetry breaking can be understood as the manifestation of a rough, glassy free energy landscape in abstract replica space~\cite{mezard1987spin}. In real space and for a given disorder realization, spin glass lattice models are characterized by an abundance of free energy minima, distinguished by their magnetization patterns. It has been proven for the classical Sherrington-Kirkpatrick model that these minima are organized into a complex hierarchical structure of large valleys, fractured into smaller and smaller valleys, see Fig.~\ref{fig_free_energy_landscape}a. Each valley is associated with a length scale, such that it contains states that share the same coarse grained magnetization pattern on this scale. Smaller valleys correspond to less coarse graining, i.e., host states that share the same magnetization pattern down to shorter scales and are therefore more correlated. This real space structure is also reflected in abstract replica space after disorder averaging. The replicas can be arranged into a hierarchical tree according to their overlaps, see Fig.~\ref{fig_free_energy_landscape}b. The leafs are labelled with the replicas, and the branching of the tree encodes a hierarchy of replica overlaps $Q_0\leq Q_1\leq Q_2\leq...$, such that the overlap between two replicas is $Q_{i}$ if the smallest branch they share starts at level $i$. These levels with a given overlap $Q_i$ are the replica  manifestations of the coarse graining on a given scale in real space, with the branches reflecting the valleys characterized by their shared coarse grained magnetization. 

The hierarchical tree sketched in Fig.~\ref{fig_free_energy_landscape}b results in a nested block diagonal structure in the Parisi overlap matrix $Q_{ab}$, with the blocks at level $m$ labelled by the overlap $Q_m$, see Fig.~\ref{fig_free_energy_landscape}c. Full replica symmetry breaking occurs when  this hierarchy of nested blocks is inifinite, i.e., each block is further partitioned into smaller and smaller blocks containing replicas with increasing overlaps. We note that  this Parisi matrix structure can also be understood  as a symmetry breaking in the permutation group of the $n$ replicas, $\mathcal{S}_n$. The replica symmetric solution preserving $\mathcal{S}_n$ corresponds to a tree with a single branch $Q_0\equiv Q_{RS}$. Inserting an additional level with $n/m$ branches, each containing $m$ leafs, i.e., adding $m\times m$ blocks along the diagonal of the Parisi matrix,  reduces the symmetry group as
\begin{equation*}
    \mathcal{S}_n\rightarrow\mathcal{S}_m^{\otimes n/m}\otimes \mathcal{S}_{n/m}.
\end{equation*}
This symmetry group consists of the permutation of replicas within a single block,  $\mathcal{S}_m$, for each of the $n/m$ blocks,  and the permutation of the full blocks, $\mathcal{S}_{n/m}$. 
In the replica limit $n\to 0$, the new symmetry group $\mathcal{S}_m^{\otimes n/m}\otimes \mathcal{S}_{n/m}$ contains the original symmetry $S_0$ as a subgroup, therefore, the same pattern of permutation symmetry breaking can be repeated infinitely many times, leading to the nested blockdiagonal structure of the Parisi matrix. 

Full replica symmetry breaking in the quantum model ~\eqref{eq:Srep}-\eqref{eq:selfconsistent} is taken into account by following the same steps as in the classical Sherrington-Kirkpatrick spin glass model. For completeness, we briefly review this derivation below. The argument relies on constructing the effective action $S_m(y)$, and an effective field distribution $P_m(y)$, for the   $m\times m$ Parisi blocks in replica space. The resulting effective model governs all physical properties at scale $m$, i.e., allows to determine all spin correlations between replicas inside the block. One then derives so called flow equations, describing how these functions change with the scale $m$, and thereby allowing to determine all physical properties from solving the effective action at the single scale $m=1$, i.e., at the replica diagonal point. We note that this calculation can be understood as a manifestation of a renormalization group procedure in real space, involving coarse graining on larger and larger scales,  reformulated in abstract replica space.

First, one introduces a ``restricted" action $S_m(y)$ and a corresponding free energy density $\phi_m(y)$, defined on a single $m\times m$ block of the Parisi matrix with $1\leq m\leq n$,
\begin{align*}
    &S_m(y)= -\dfrac{J^2}{2}\sum_{a, b=1}^m \left(Q_{ab}-Q_m\right)\int_\tau\int_{\tau^\prime}\sigma^z_{a\,\tau}\sigma^z_{b\,\tau^\prime}+\\
    &\;\sum_{a=1}^m\left[\int_\tau\left(y\,\sigma^z_{a\tau}+h_T\,\sigma^x_{a\tau}\right)-\dfrac{J^2}{2}\int_\tau\int_{\tau^\prime}\tilde\chi(\tau-\tau^\prime)\sigma^z_{a\,\tau}\sigma^z_{a\,\tau^\prime}\right],
\end{align*}
and
\begin{equation*}
    e^{\beta\, m\, \phi_m(y)}=\int\mathcal{D}\sigma^z\, e^{-S_m(y)},
\end{equation*}
where $\tilde\chi(\tau)\equiv \chi(\tau)-Q_{aa}$, with $Q_{aa}=\lim_{m\to 1} Q_m$.  Here, the action $S_m(y)$  was obtained by restricting the replica sums in $S_{\rm rep}$ to indices $1\leq a\leq m$, and by eliminating the ``background" coupling $Q_m$ by a Hubbard-Stratonovich transformation, taken into account through the random field $y$, similarly to the treatment of the replica symmetric action in Sec.~\ref{subsec:rs}. As a result, the replicas within the $m\times m$ Parisi block become decoupled, unless they share the same $(m-\Delta m)\times (m-\Delta m)$ block at the next level $m-\Delta m$. In terms of the free energy landscape, this procedure can be interpreted as considering the partition function of a single valley at level $m$. 

Importantly, the replica diagonal action, $S_1(y)$, has the same structure as the replica symmetric result Eq.~\eqref{eq:S_y}.  The corresponding free energy density, $\phi_1(y)$, is accessible by a continuous-time Monte Carlo approach, described below in Sec.~\ref{sec:ctqmc}. All other free energy densities $\phi_m(y)$ can then be obtained by deriving a recursion relation, expressing $\phi_{m+\Delta m}(y)$ in terms of $\phi_m(y)$, see Appendix~\ref{app:flow} for details. Finally, in the replica limit $n\to 0$, the label $m$ transforms to a continuous variable $x\in [0,1]$, $Q_m$ evolves into a monotonously increasing function $Q(x)$, and the recursion relation becomes a partial differential equation for the function $\phi(x,y)$,
\begin{equation}
    \partial_x\phi(x,y)=-\dfrac{J^2}{2}\dfrac{dQ}{dx}\left\lbrace \partial_y^{\,2}\,\phi(x,y)+\beta x\left(\partial_y\phi(x,y)\right)^2\right\rbrace.
\label{eq:flow_phi}
\end{equation}
This flow equation describes the evolution of the free energy density of a valley at scale $x$ in random field $y$, upon changing the scale $x$. Therefore, it allows us to determine the free energy of the full system from the replica diagonal boundary condition $\phi_1(y)$.

The second ingredient for obtaining the equations governing full replica symmetry breaking is introducing a scale dependent distribution function $P_m(y)$, encoding the distribution of the random magnetic field $y$ within a valley at scale $m$, appearing in the reduced action $S_m(y)$. This distribution incorporates the mean field renormalization of the bare disorder by the interactions between the Parisi block in question, and the rest of the system. It is determined by imposing the condition that any spin correlator within the $m\times m$ Pauli block, $\mathcal{O}_m$ can be obtained by evaluating it with respect to $S_m(y)$, followed by a disorder average over $y$ according to $P_m(y)$,
\begin{equation*}
    \langle \mathcal{O}_m\rangle_{S_{\rm rep}}=\int dy\, P_m(y)\,\langle \mathcal{O}_m\rangle_{S_{m}(y)}, \text { for $\mathcal{O}_m$ arbitrary}.
\end{equation*}

We note that at the boundary $m=n$, $S_n(y)$ follows from the replicated action ~\eqref{eq:Srep} by a single Hubbard-Stratonovich transformation over a uniform replica offdiagonal coupling $Q_n\equiv Q_0$. Therefore, at the boundary $m=n$, $P_n(y)$ preserves the Gaussian form of the bare disorder, with a variance renormalized by interactions, $h_z^2\rightarrow h_z^2+J^2 Q_0$, similarly to the replica symmetric result ~\eqref{eq:PRS}. The distribution at an arbitrary scale $m$ can then be determined by following the strategy applied for the free energy density, and deriving a recurrence relation between $P_m(y)$ and $P_{m+\Delta m}(y)$, see Appendix ~\ref{app:flow} for details. Performing the replica limit and switching to the continuous variable $x$ yields the flow equation,
\begin{align}\label{eq:flow_P}
    &\partial_x P(x,y)=\nonumber\\
    &\quad\dfrac{J^2}{2}\dfrac{dQ}{dx}\left\lbrace \partial_y^{\,2} P(x,y)-2\beta x\,\partial_y\left[P(x,y)\partial_y\phi(y,x)\right]\right\rbrace.
\end{align}

To summarize the result of these rather technical considerations, the solution in the presence of full replica symmetry breaking can be obtained by solving the following self-consistency problem iteratively. We first initialize the function $Q^{[0]}(x)$, with finite derivative to allow full replica symmetry breaking, and a dynamical spin corretalor $\tilde\chi^{[0]}(\tau)$. Then, in each step of the iteration, we substitute $\tilde\chi^{[i]}(\tau)$ into the replica diagonal action ~\eqref{eq:S_y}, and  solve it with CTQMC for $\langle\sigma^z\rangle_{S(y)}$ and $\tilde{\chi}_y(\tau)$ on a fine grid in $y$. We then proceed by noting that $\langle\sigma^z\rangle_{S(y)}$ is the derivative of the replica diagonal free energy density,
\begin{equation*}
    \langle\sigma^z\rangle_{S(y)}=\partial_y\phi_1(y),
\end{equation*}
and rewrite the flow equation ~\eqref{eq:flow_phi} in terms of a scale dependent magnetization $m(x,y)=\partial_y\phi(x,y)$,
\begin{equation*}
    \partial_x m(x,y)=-\dfrac{J^2}{2}\dfrac{dQ}{dx}\left\lbrace \partial_y^{\,2}\,m(x,y)+\beta x\partial_y\left[ m(x,y)^2\right]\right\rbrace.
\end{equation*}
We get the magnetization at all scales by numerically solving this differential equation with boundary condition $m(1,y)=\langle\sigma^z\rangle_{S(y)}$, using the function $dQ^{[i]}/dx$. The next step is to set the boundary condition $P(0,y)$ to a Gaussian distribution with variance $h_z^2+J^2 Q^{[i]}(0)$, and to solve Eq.~\eqref{eq:flow_P} by substituting $m(x,y)$ and $dQ^{[i]}/dx$ into the right hand side. Finally, we update all parameters for the next iteration step. According to the definition of $P(x,y)$, a replica diagonal expectation value is obtained by evaluating it with respect to $S(y)$, and averaging over $y$ according to the distribution $P(1,y)$,
\begin{align*}
    \tilde\chi^{[i+1]}(\tau)=\int dy\, P(1,y)\,\tilde{\chi}_y(\tau),
\end{align*}
whereas the overlap $Q(x)$ follows from the ``restricted" action at scale $x$, and the corresponding distribution $P(x,y)$,
\begin{align*}
    Q^{[i+1]}(x)=\int dy\, P(x,y)\,m(x,y)^2.
\end{align*} 
In this last relation, we used that a pair of replicas in the same Parisi block at scale $x$, but in different blocks at all larger scales, is decoupled in the effective action at scale $x$, therefore, their overlap is simply the square of the average magnetization $m(x,y)^2$. This equation can be rewritten in a more convenient form by taking the derivative with respect to $x$, and using the flow equations,
\begin{align}\label{eq:dQ_update}
   \dfrac{dQ^{[i+1]}}{dx}=\dfrac{dQ^{[i]}}{dx}J^2\int dy\, P(x,y)\,\left[\partial_ y m(x,y)\right]^2.
\end{align}
These updating formulas close the iteration step, and the procedure can be repeated until convergence.

According to Eq.~\eqref{eq:dQ_update}, the converged solution has to to satisfy
\begin{align}\label{eq:marginality}
   1=J^2\int dy\, P(x,y)\,\chi_{\rm suscep}(x,y)^2,
\end{align}
where we introduced the scale dependent susceptibility $\chi_{\rm suscep}(x,y)=\partial_ y m(x,y)$. This relation encodes the so-called marginal stability of the glassy phase, ensuring that solution with full replica symmetry breaking remains marginally stable against perturbations at all scales.

We close this section by noting that despite the technical difficulties arising in the presence of full replica symmetry breaking, the structure of the resulting equations can be well understood based on the replica symmetric case, as well as the cavity approach. The replica diagonal action ~\eqref{eq:S_y}, also obtained from more intuitive cavity considerations, still governs the physical properties. The only subtlety is the more complex renormalization of effective disorder distribution $P(y)$ due to the complex interplay of interaction terms, completely erasing the Gaussian structure of the bare disorder. These effects are systematically incorporated into the replica approach through the flow equations ~\eqref{eq:flow_phi} and ~\eqref{eq:flow_P}.

\section{Continuous-time quantum Monte Carlo approach}\label{sec:ctqmc}

As discussed in the previous section, for the complete solution of the model we need to compute the quantities $\langle \sigma^{z} \rangle_{y}$ and $\tilde{\chi}_{y}(\tau)$ for a large set of the effective magnetic fields $y$, by performing CTQMC calculation with the effective local action~\eqref{eq:S_y}.
We use an $h_{T}$-expansion CTQMC algorithm, well suited for incorporating  retarded interactions in the action formalism\cite{werner_EDMFT2013, werner_dynamical_screening2010}. We outline the main ingredients of this method below, with the technical details left to Appendix ~\ref{app-details-of-CTQMC}.

\begin{figure}[t]
\centering
\includegraphics[width=0.8\columnwidth]{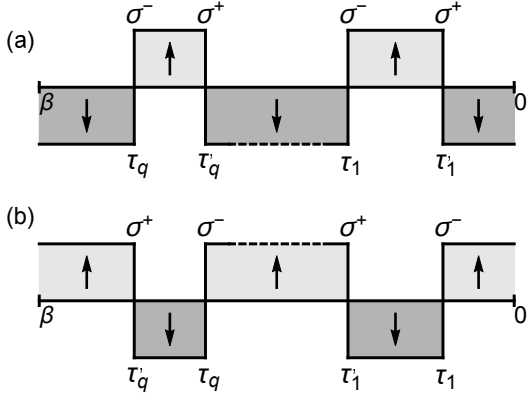}
\caption{ CTQMC approach for solving the  local effective action. 
Configurations with operator sequences $ \hat{\sigma}^{+}_{\tau_{1}^{\prime}}\hat{\sigma}^{-}_{\tau_{1}} ...  \hat{\sigma}^{+}_{\tau_{q}^{\prime}}\hat{\sigma}^{-}_{\tau_{q}}$ (a) and $ \hat{\sigma}^{-}_{\tau_{1}}\hat{\sigma}^{+}_{\tau^{\prime}_{1}} ...  \hat{\sigma}^{-}_{\tau_{q}}\hat{\sigma}^{+}_{\tau^{\prime}_{q}}$ (b) contributing to the partition function in the $h_T$-expansion, visualized in the segment picture.
}\label{Fig-segments}
\end{figure}

In this CTQMC approach, we expand the partition function $Z_{y}={\rm Tr}\,{\rm e}^{-S(y)}$ in terms of the transverse magnetic field $h_{T}$ as 
\begin{eqnarray}
Z_{y}  &=&   \sum_{q=0}^\infty \left( \frac{1}{q!}\right)^2 \left( h_{T} \right)^{2q}\, 
  \prod_{i=1}^q\,\int_{\tau_i} \int_{\tau_i^{\prime}} {\rm Tr}\, \left[  {\rm e}^{-(S_{z}(y) + S_{\tilde{\chi}})}\times\nonumber\right. \\ 
 && \quad\quad\left.\left( \hat{\sigma}^{+}_{\tau^{\prime}_{1}} \hat{\sigma}^{-}_{\tau_{1}} ... \hat{\sigma}^{+}_{\tau^{\prime}_{q}} \hat{\sigma}^{-}_{\tau_{q}}   + \hat{\sigma}^{-}_{\tau_{1}}\hat{\sigma}^{+}_{\tau^{\prime}_{1}}  ...  \hat{\sigma}^{-}_{\tau_{q}} \hat{\sigma}^{+}_{\tau^{\prime}_{q}} \right)\right],
 \label{Eq-partition-function-expansion-SpinGlass}
\end{eqnarray}
and sample the sum of multiple integrals stochastically. Here we defined  the actions
\begin{equation}\label{eq:Szy}
    S_{z}(y) \equiv y\int_{\tau} \sigma^{z}_{\tau},
\end{equation}
and
\begin{equation}\label{eq:Schi}
  S_{\tilde{\chi}} \equiv -\dfrac{J^2}{2} \int_{\tau} \int_{\tau^{\prime}} \tilde{\chi}(\tau-\tau^{\prime}) \sigma^{z}_{\tau} \sigma^{z}_{\tau^{\prime}},  
\end{equation}
only depending on the $z$ component of the spin. We also used that due to $(\hat{\sigma}^{+})^2=(\hat{\sigma}^{-})^2=0$, 
only the operator sequences $\hat{\sigma}^{-}\hat{\sigma}^{+}...\hat{\sigma}^{+}$ and $\hat{\sigma}^{+}\hat{\sigma}^{-}...\hat{\sigma}^{-}$ contribute to the partition function in the above expansion.

\begin{figure*}[t]
\centering
\includegraphics[width=0.31\textwidth]{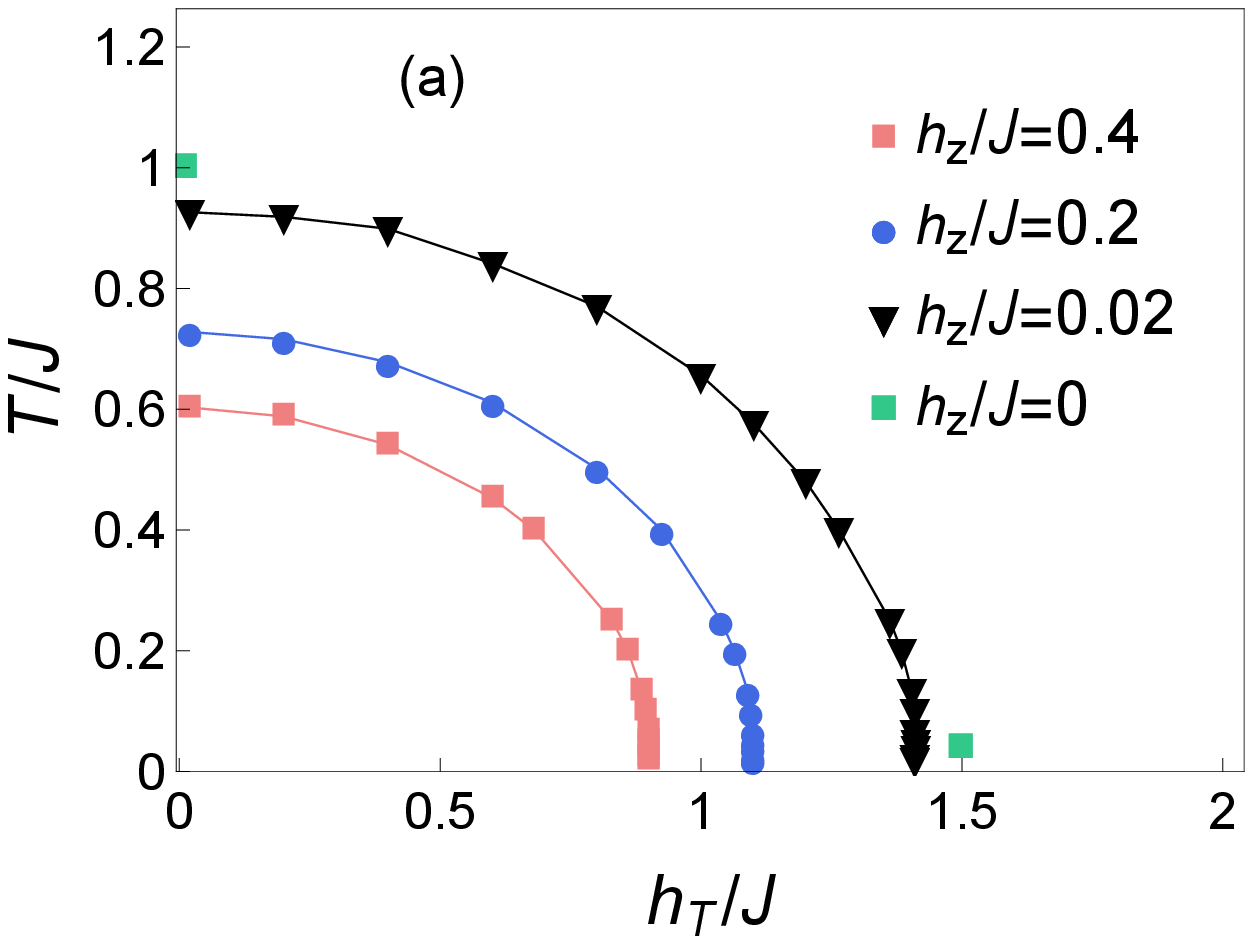}\hspace*{0.5cm}
\includegraphics[width=0.31\textwidth]{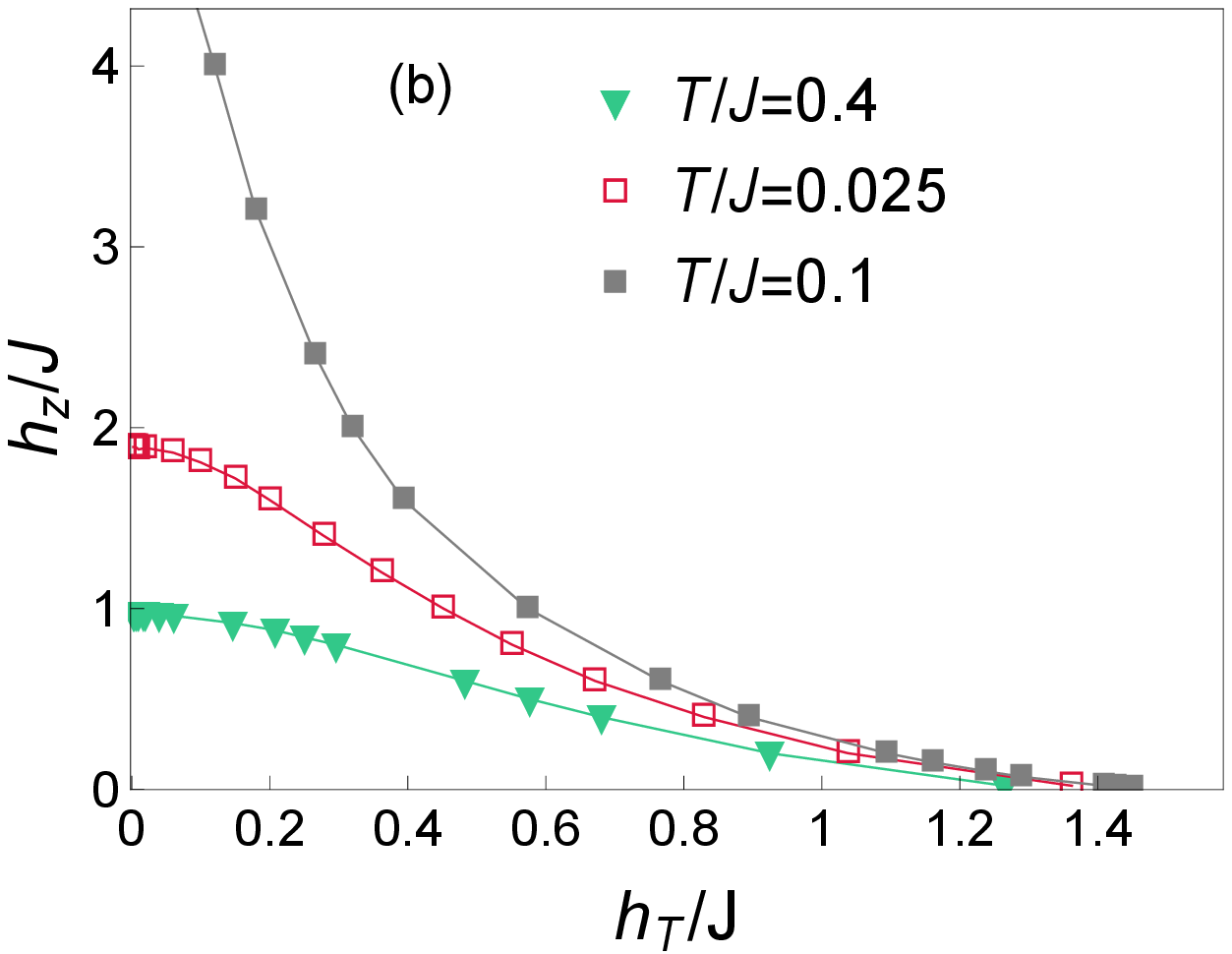}\hspace*{0.5cm}
\includegraphics[width=0.31\textwidth]{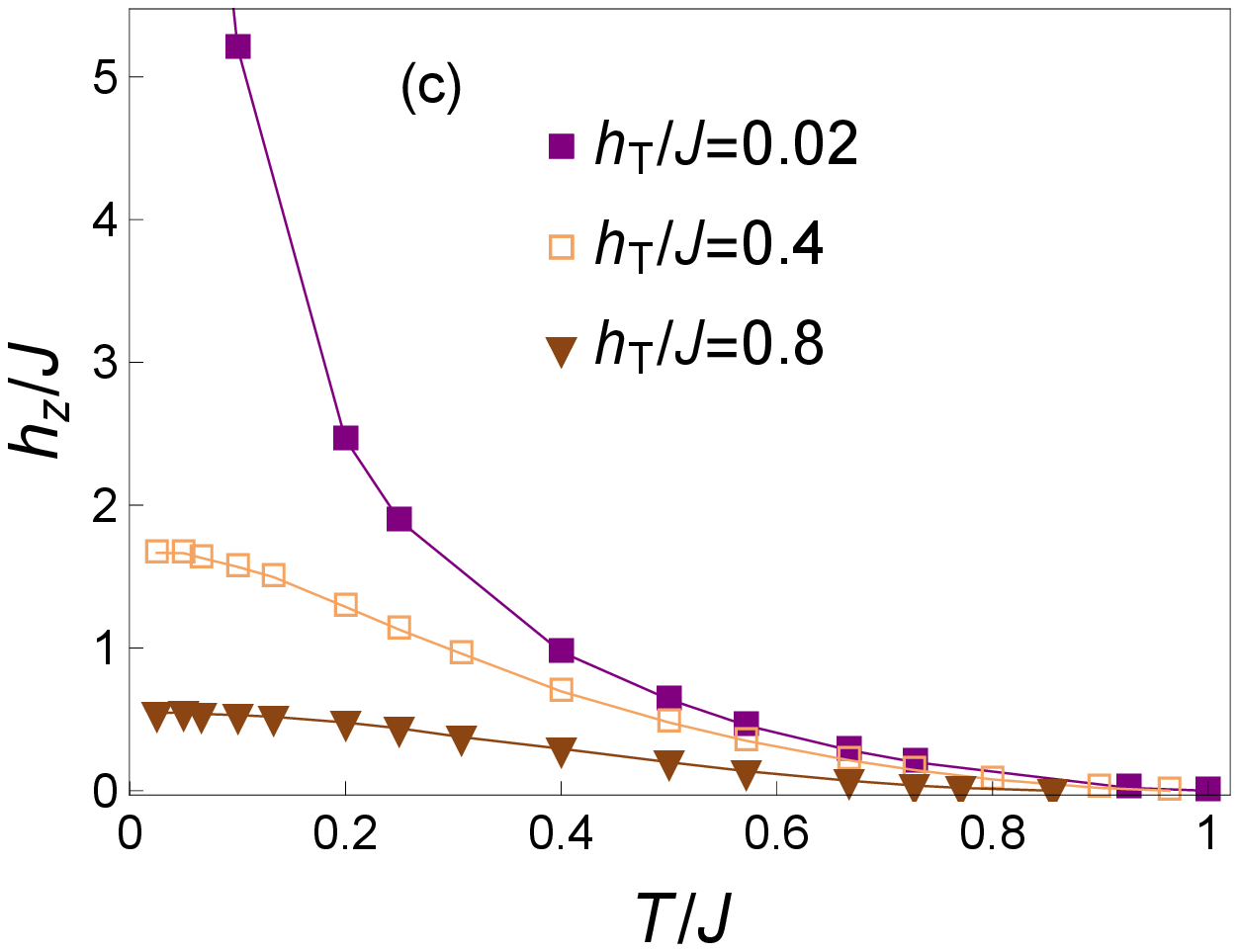}
\caption{Two-dimensional cuts of the phase diagram.
(a) Cuts in $h_{\rm T}/J$-$T/J$ plane shown  for various values of $h_{z}/J$. Glass can be melted by thermal or quantum fluctuations, and the glassy region shrinks with increasing $h_z$. 
(b) Cuts in $h_{\rm T}/J$-$h_{z}/J$ plane for different temperatures $T/J$. 
(c) Phase diagram in $T/J$-$h_{z}/J$ plane for different transeverse fields $h_{\rm T}/J$. 
}
\label{fig-phase-diagrams}
\end{figure*}

We can rewrite this expression for the partition function as an integral over configurations,
\begin{equation*}
    Z_{y}  =  \int {\cal D}( {\boldsymbol{\tau}_{q}})\, w({\boldsymbol{\tau}_{q}}),
\end{equation*}
where a configuration ${\boldsymbol{\tau}_{q}} = \{ \tau^{\prime}_{1}, \tau_{1}, ... \tau^{\prime}_{q-1},  \tau_{q-1}, \tau^{\prime}_{q}, \tau_{q} \}$ for $\sigma^{z}_{\tau=0}=-1$ and ${\boldsymbol{\tau}_{q}} = \{ \tau_{1}, \tau^{\prime}_{1}, ... \tau_{q-1},  \tau^{\prime}_{q-1}, \tau_{q}, \tau^{\prime}_{q} \}$ for $\sigma^{z}_{\tau=0}=1$ is a set of imaginary times at which the operations $\hat{\sigma}^{-}_{\tau_{k}}$ and $\hat{\sigma}^{+}_{\tau^{\prime}_{k}}$ occur,
and $\int {\cal D}({\boldsymbol{\tau}_{q}}) = \sum_{q} \prod_{i=1}^q\int_{\tau_1} \int_{\tau_i^{\prime}}$, with the  imaginary times ordered either as $\beta > \tau_{q}^{\prime} > \tau_{q} > ... > \tau_{1}^{\prime} > \tau_{1} >0 $ or as $\beta > \tau_{q} > \tau_{q}^{\prime} > ... > \tau_{1} > \tau_{1}^{\prime} >0 $, depending on the value of $\sigma^z$ at $\tau=0$. The operators $\hat{\sigma}^{-}_{\tau_{k}}$ and $\hat{\sigma}^{+}_{\tau^{\prime}_{k}}$  flip  $\sigma^{z}$ as $1 \rightarrow -1$ at the imaginary times $\tau_{k}$ and $-1 \rightarrow 1$ at the imaginary times $\tau^{\prime}_{k}$, respectively, producing a sequence of alternating signs for $\sigma^{z}$, periodic in $\beta$ due to the Tr operation. These sequences are conveniently represented by segments as illustrated in Fig.~\ref{Fig-segments}.
Performing the trace calculation in Eq.~(\ref{Eq-partition-function-expansion-SpinGlass}),  we obtain a total weight
\begin{equation}\label{eq:weights}
  w({\boldsymbol{\tau}_{q}}) = h_{T}^{2q}\, w_{z}({\boldsymbol{\tau}_{q}},y)\, w_{\tilde{\chi}}({\boldsymbol{\tau}_{q}}),  
\end{equation}
 with the contributions $w_{z}$ and $w_{\tilde{\chi}}$ stemming from the effective field $y$ through the action $S_z(y)$, and the interaction term through $S_{\tilde\chi}$, respectively. The explicit formulas for these weight factors, as well as their derivation,  are presented in Appendix~\ref{app:ctqmc_weights}.

The observables $\langle \sigma^{z} \rangle_{y}$ and $\tilde{\chi}_{y}(\tau)$ can be evaluated in the CTQMC method by sampling the segment configurations stochastically, according their weight $w({\boldsymbol{\tau}_{q}})$, and evaluating the contribution of each configuration to these operator expectation values. We perform this random sampling via a Metropolis algorithm, described in Appendix ~\ref{app:Metropolis}. More details on the calculation of the observables are provided in Appendix ~\ref{app:observables}.

The iterative solution of the self-consistency problem under replica symmetry and in the presence of full replica symmetry breaking goes as it was outlined in Sections~\ref{subsec:rs} and \ref{subsec:rsb}, respectively.

\section{Numerical results}\label{sec:numerics}

The theoretical framework described in Sec. ~\ref{subsec:replica} and the application of the quantum Monte Carlo algorithm presented in Sec.~\ref{sec:ctqmc}
allows us to obtain the numerically exact solution of the quantum spin glass model ~\eqref{eq:hamiltonian}. In this section, we present our numerical results concerning the phase diagram, and also the properties of the spin glass phase including the order parameter, distribution of the effective magnetic fields, as well as the dynamics.

\subsection{Phase diagram} 

We calculate the solution within the replica symmetric paramagnetic phase by solving the self-consistency equations ~\eqref{eq:S_y}-\eqref{eq:chiRS}. To remain stable  against full replica symmetry breaking, this solution has to satisfy the following stability criterion,
\begin{align}\label{eq:RS_stab}
   1\geq J^2\int dy\, P_{RS}(y)\, \tilde\chi_{\rm st}(y)^{2},
\end{align}
with the static susceptibility $\tilde\chi_{\rm st}(y)=\int_\tau \tilde\chi_{y}(\tau)$. This stability condition follows immediately from comparing to Eq.~\eqref{eq:dQ_update}, describing how a small symmetry breaking term evolves under iteration, or, alternatively, it can be derived directly by inspecting how the free energy density changes as a result of a symmetry breaking perturbation.

We determine the full phase diagram of the model in terms of the parameters $T/J$, $h_T/J$ and $h_z/J$, by finding the points in parameter space where the replica symmetric solution becomes marginally stable, i.e., Eq.~\eqref{eq:RS_stab} is satisfied with as an equality. 

The summary of our results are shown in Fig.~\ref{fig-3D-phase-diagram}, displaying a spin glass phase with full replica symmetry breaking at low enough temperatures and  transverse fields. This glassy phase is eventually melted by thermal and quantum fluctuations upon increasing $T$ and $h_T$. Importantly, besides these effects, a strong enough onsite disorder $h_z$ also melts the glass, through inducing a trivial state where each spin aligns independently with the strong local field $h_i$. These effects are further illustrated in Fig.~\ref{fig-phase-diagrams}, showing two-dimensional cuts of the full phase diagram in the planes $h_T-T$ (a), $h_T-h_z$ (b) and $T-h_z$ (c).
 
Taking directly $h_{z}=0$ in our calculations is computationally hard around the phase transition since the distribution $P_{RS}(y)$  in Eq.~(\ref{eq:RS_stab}) becomes a Dirac delta at the transition where $Q_{RS}=0$ for $h_{z}=0$.
Using the scaling property of the order parameter $Q_{RS}$ that it vanishes linearly in the vicinity of the critical value of $h_{T}$\cite{young2017stability}, we extrapolate our $Q_{RS}$ data to $h_{z}=0$ and obtain $h_T/J=1.5$ for the critical transverse field close to the zero temperature limit, at $T/J=0.04$, in good agreement with previous estimates in the literature \cite{miller_huse1993, rozenberg_grempel1998, young2017stability}.  
In the classical limit, $h_{T}=0$, with similar procedure we obtain the known result $T/J=1$ for the critical temperature.
These results are included in the {\sl right panel} of Fig.~\ref{fig-phase-diagrams}

\subsection{Distribution of the effective magnetic fields}

\begin{figure*}[t!]
\centering
\includegraphics[width=0.37\textwidth]{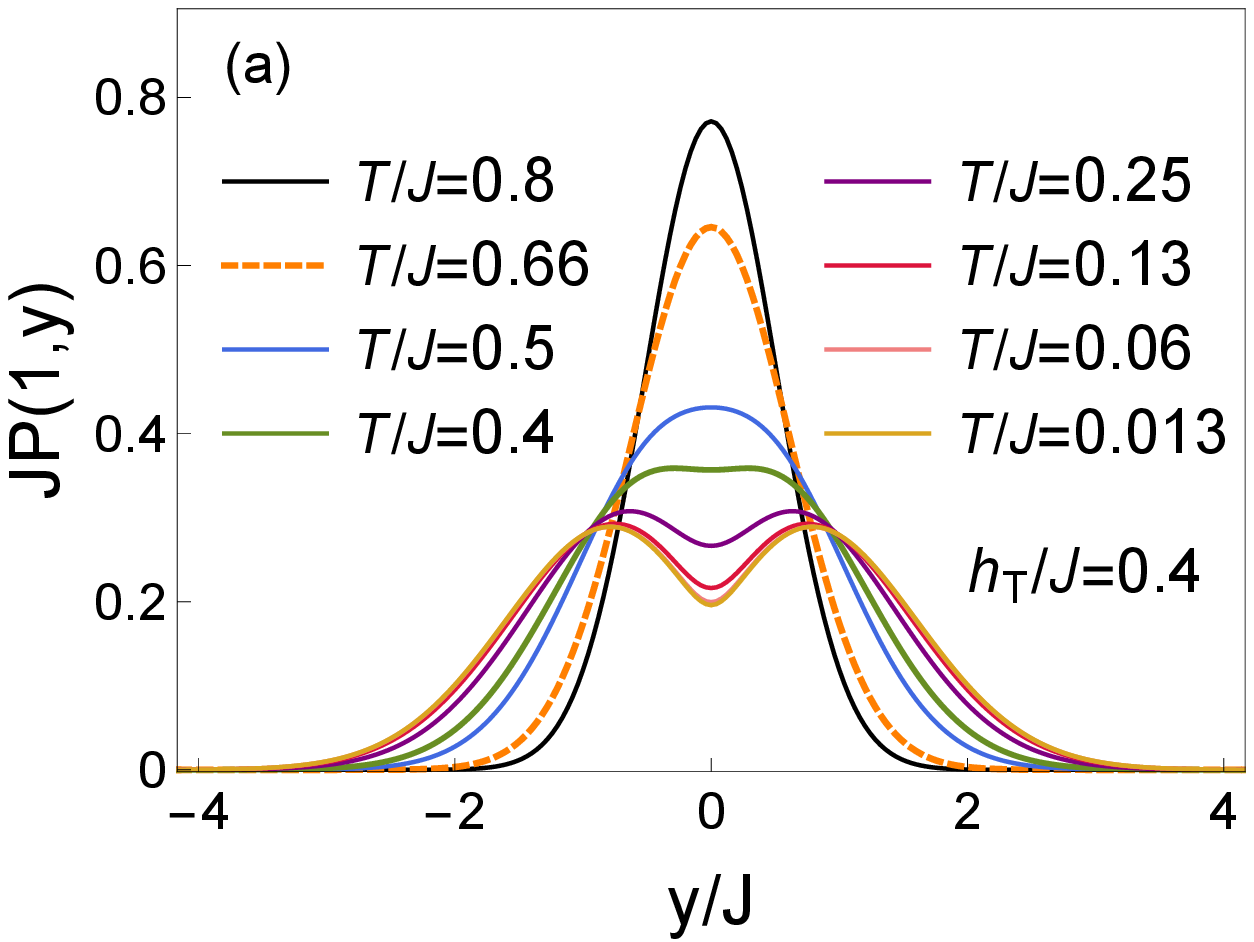}\hspace*{0.8cm}
\includegraphics[width=0.37\textwidth]{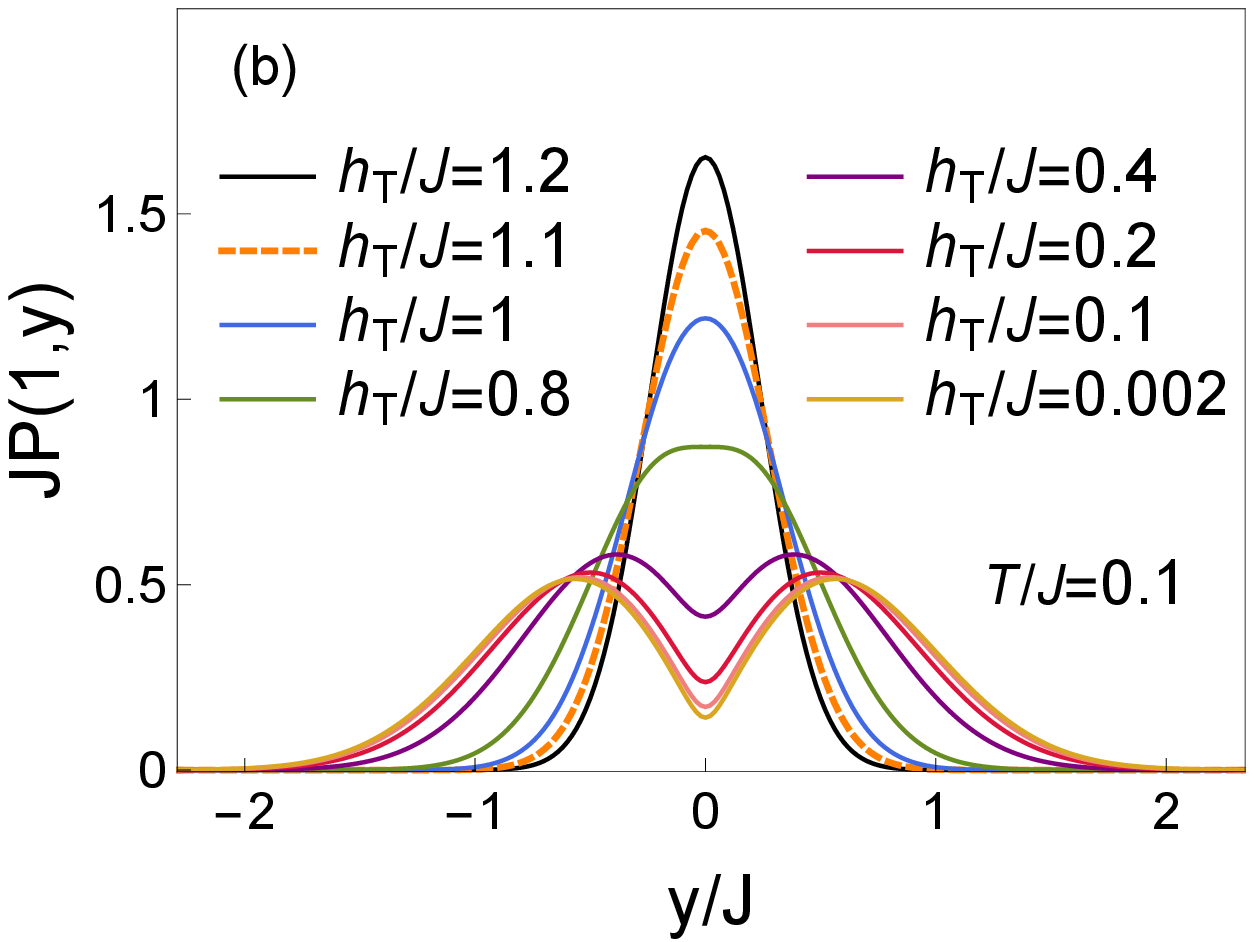}
\caption{Renormalized distribution of local $z$ magnetic field. Evolution of $P(1,y)$ distribution with decreasing temperature at $h_{\rm T}/J=0.4$ fixed (a), and with decreasing transverse field $h_{\rm T}$ at $T/J=0.1$ (b). Distribution remains Gaussian in the paramagnetic phase, while a pseudo-gap opens up in the spin glass phase, converging towards a universal scaling form $P(1,y)\sim |y|/J^2$ for small fields $y$. Distributions at the phase boundary shown by orange dashed lines. 
We used  $h_z/J$=0.2.}
\label{fig_Pxy_distribution}
\end{figure*}

Having obtained the complete phase diagram, we now turn to the properties of the spin glass phase, by applying the iterative procedure described in Sec.~\ref{subsec:rsb}. As already discussed there, all correlations within a single replica are still governed by the replica diagonal action ~\eqref{eq:S_y}, but the distribution of the random magnetic field $y$ appearing in this action is renormalized by the interactions compared to the Gaussian bare disorder. This renormalization  keeps the Gaussian shape intact in the paramagnetic phase, only changing the variance according to Eq.~\eqref{eq:PRS}. The renormalization is more complex  in the glassy phase, manifesting in the changing shape of  $P(1,y)$ as we go deeper into the glassy phase.

The evolution of $P(1,y)$ across the phase boundary and within the glassy phase is shown in Fig.~(\ref{fig_Pxy_distribution}), displaying the deformation of this distribution with decreasing thermal (a) or quantum (b) fluctuations. A dashed line denotes the Gaussian shape  at the phase boundary. Upon entering the glassy phase, the distribution develops a pseudo-gap structure, i.e., the probability of a small fields $y$ is strongly suppressed.

Such a pseudo-gap formation is a characteristic feature of glassiness, and gives rise to a universal scaling deep within the spin glass phase, $P(1,y)\sim |y|/J^2$ for fields $y$ small enough. Importantly, this universal result only depends on the interaction strength $J$, showing that the glass transition is a structural phase transition, governed by the complicated interplay of frustrated interactions. We note that the universal form of the pseudogap can be understood based on simple, classical stability arguments, by inspecting the stability of the state against flipping pairs of spins.

\subsection{Order parameter and the overlap distribution}

\begin{figure}[t]
\centering
\includegraphics[width=0.85\columnwidth]{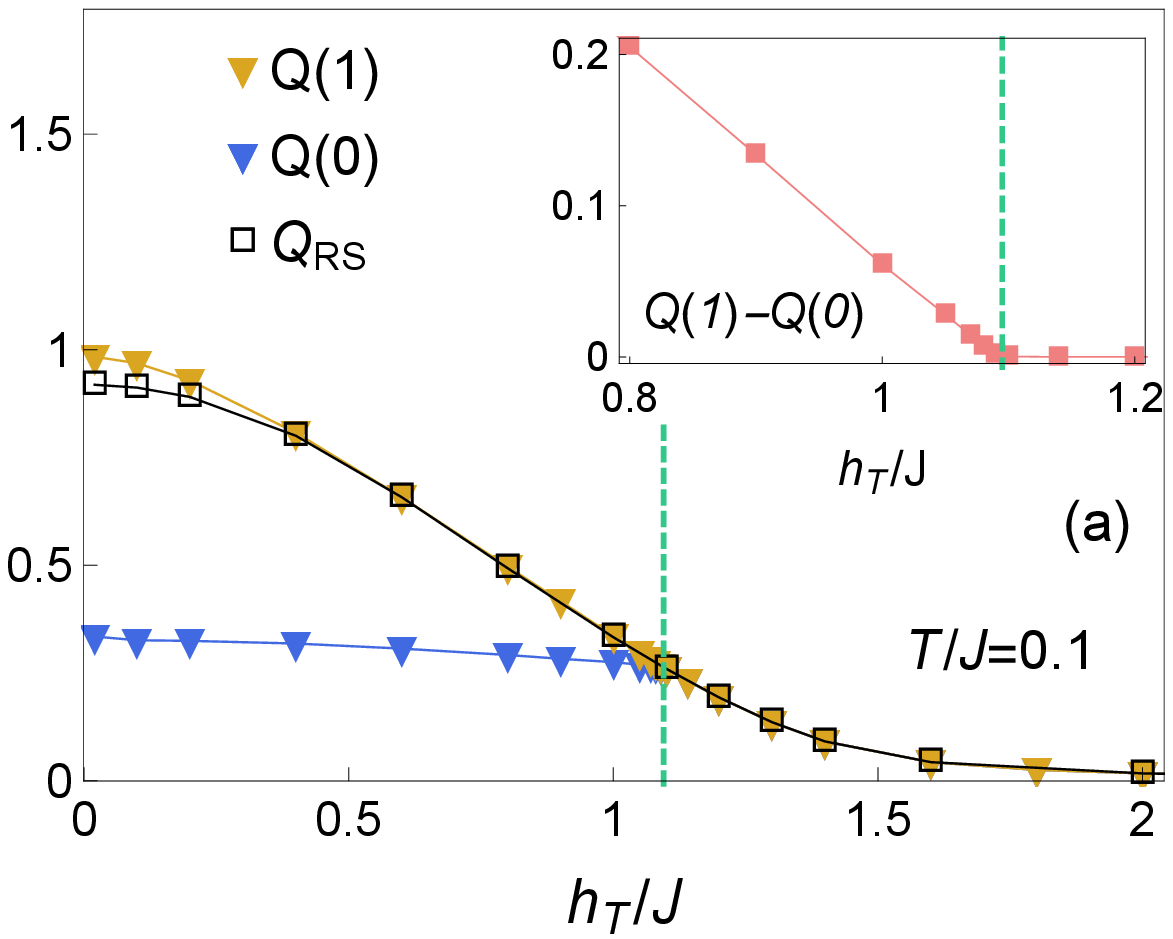}\vspace*{0.5cm}
\includegraphics[width=0.9\columnwidth]{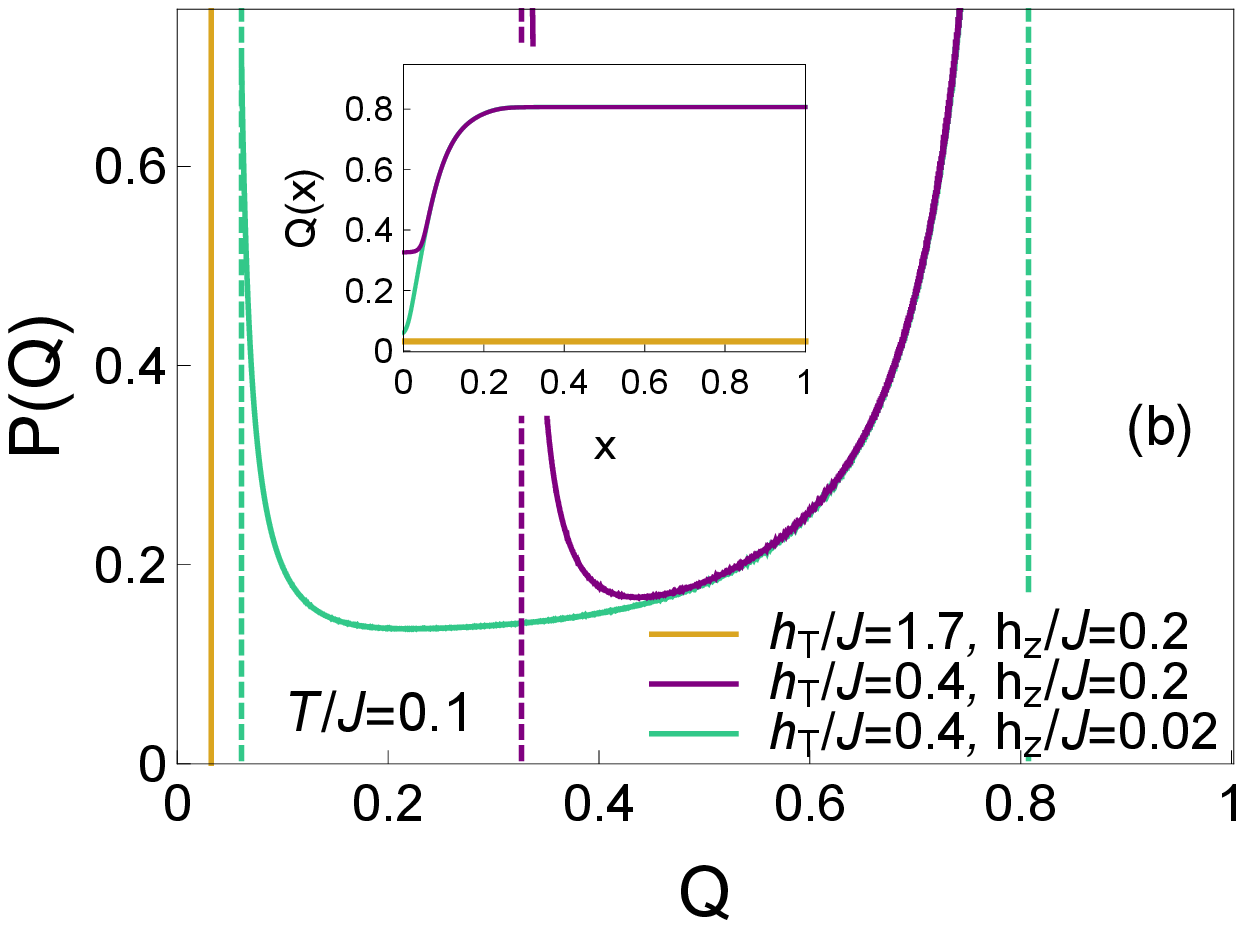}
\caption{Order parameter and overlap distribution in the spin glass phase. (a) The overlaps $Q(1)$, $Q(0)$, and $Q_{\rm RS}$ as a function of transverse field $h_T$, across the spin glass phase boundary (green dashed line). The order parameter $Q(1)-Q(0)$  is finite in the glassy phase, but vanishes in the paramagnetic region. Inset: critical scaling of $Q(1)-Q(0)$, vanishing linearly as $h_T$ approaches the critical point from below. (b) Overlap distribution $P(Q)$ in the paramagnetic phase for $h_{\rm T}/J$=1.7 and $h_z/J$=0.2 (orange), and for two parameter sets within the spin glass phase, $h_{\rm T}/J$=0.4, with  $h_z/J$=0.2 (purple) and $h_z/J$=0.02 (green). Distribution is a single Dirac-delta in the paramagnetic phase, developing a non-trivial continuous structure in a broadening range $[Q(0),Q(1)]$ in the glassy phase. The minimal overlap $Q(0)$ goes to zero in the limit of $h_{z}$=0. Inset: Overlap functions $Q(x)$, corresponding to the distributions in the main panel.
}
\label{fig_PQ-Q_P_pseudogap}
\end{figure}

In the replica formalism, full replica symmetry breaking is encoded in the overlap function $Q(x)$. In the paramagnetic phase, $Q(x)\equiv Q_{RS}$, whereas the spin glass phase is characterized by a monotonous function with $Q(1)-Q(0)>0$. Therefore, the difference between maximal and minimal overlaps, $Q(1)-Q(0)$, serves as an order parameter for the transition.

The overlaps $Q(1)$ and $Q(0)$, as well as the replica symmetric solution $Q_{RS}$ are shown in Fig.~\ref{fig_PQ-Q_P_pseudogap}a as a function of transverse field $h_{T}$, across the phase boundary indicated by a vertical green dashed line. In the paramagnetic phase at large $h_T$, $Q(0)=Q(1)=Q_{\rm RS}$, while $Q(1)-Q(0)$ starts to increase upon entering the glassy phase. In this region, the replica symmetric solution corresponding to overlap $Q_{RS}$ is unstable, therefore, it becomes unphysical. The critical scaling of the order parameter $Q(1)-Q(0)$ is displayed in the inset of Fig.~\ref{fig_PQ-Q_P_pseudogap}a, showing that it vanishes linearly as $h_T$ approaches its critical value from below \cite{young2017stability}.

As discussed in Sec.~\ref{subsec:rsb}, in the original lattice system, the spin glass transition manifests in a complex free energy landscape with a hierarchy of metastable valleys, and the overlap function $Q(x)$ of the replica formalism is closely related to the properties of this rough landscape. In particular, for classical spin glasses it has been proven that $Q(x)$ encodes the possible overlaps in the spin configurations  of two metastable states $\alpha$ and $\beta$ in the same disorder, $Q_{\alpha\beta}=1/N\sum_{i=1}^{N} \langle\sigma^z_{i}\rangle_\alpha\langle\sigma^z_{i}\rangle_{\beta}$, such that $Q(x=0)\leq Q_{\alpha\beta}\leq Q(x=1)$. More precisely, $Q(x)$ contains  the following detailed information on the \textit{full distribution} of real space overlaps $Q_{\alpha\beta}$. Provided that the states $\alpha$ and $\beta$ are sampled according to their respective Boltzmann weights, the probability density function of  $Q_{\alpha\beta}$ can be obtained from the replica calculation as~\cite{mezard1987spin}
\begin{equation*}
    P(Q_{\alpha\beta}=Q)=\left.\dfrac{dx}{dQ}\right|_{x=x(Q)}.
\end{equation*}
In the replica formalism, the same distribution $P(Q)$ describes the possible overlaps between replicas $a$ and $b$, $Q_{ab}$.

We show the overlap distribution between replicas, $P(Q)$, in Fig.~\ref{fig_PQ-Q_P_pseudogap}b, for a transverse field  $h_{T}/J=1.7$ corresponding to the paramagnetic phase,
and for $h_{T}/J=0.4$ with two different $h_z$ values within the spin glass phase. 
In the paramagnetic phase, $Q(x)\equiv Q_{RS}$ gives rise to a trivial distribution consisting of a single Dirac-delta function, $P(Q)=\delta\left(Q-Q_{\rm RS}\right)$. Upon entering the spin glass phase, $P(Q)$ acquires a non-trivial structure over a finite range $Q\in [Q(0),Q(1)]$,  broadening as we go deeper into the glassy phase. In particular, $Q(0)$ approaches 0 with decreasing typical bare magnetic field $h_z$, while the maximal overlap $Q(1)$ stays close to its maximal value 1.

In the inset of Fig.~\ref{fig_PQ-Q_P_pseudogap}b, we also show the overlap functions $Q(x)$ corresponding to three distributions displayed in the main panel.

\subsection{Static and dynamical susceptibility}

\begin{figure}[t]
\centering
\includegraphics[width=0.8\columnwidth]{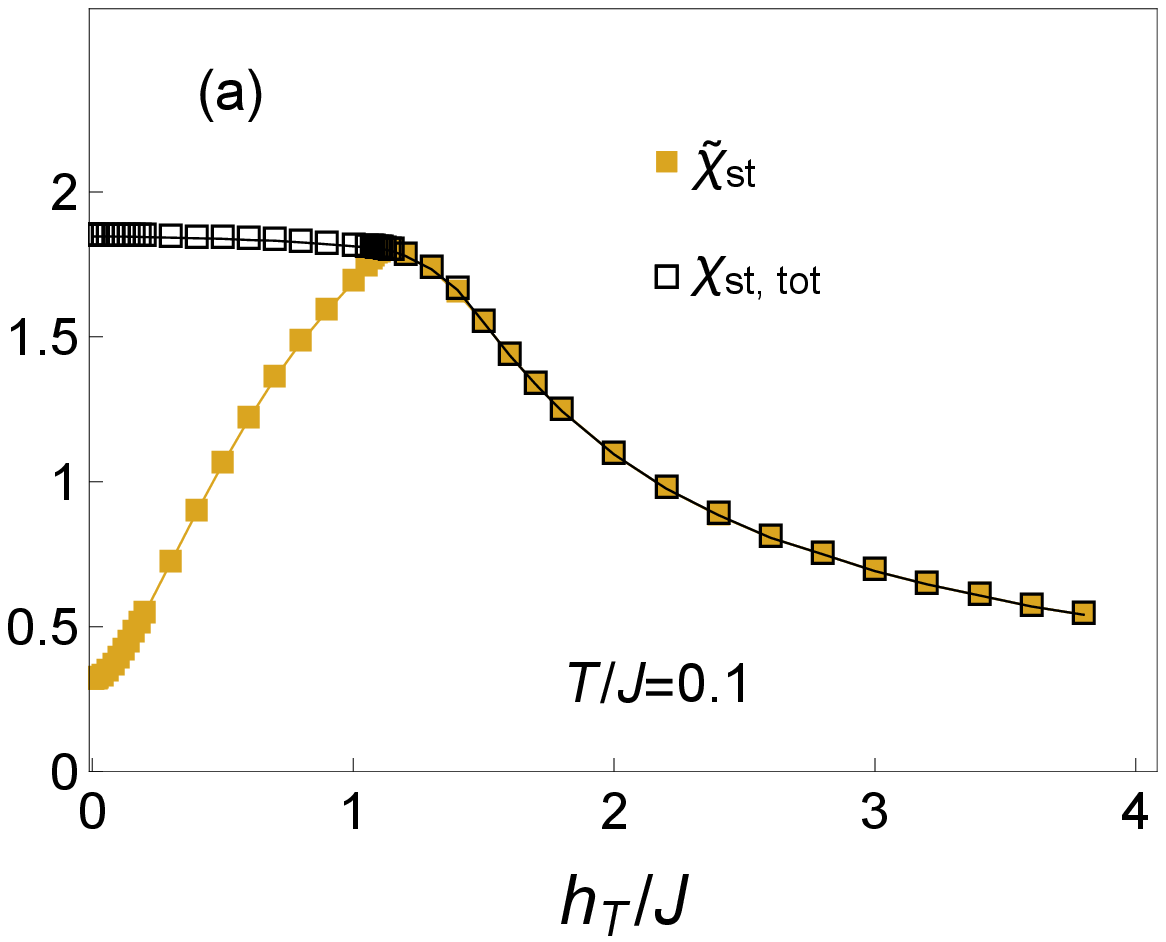}\hspace{0.5cm}
\includegraphics[width=0.8\columnwidth]{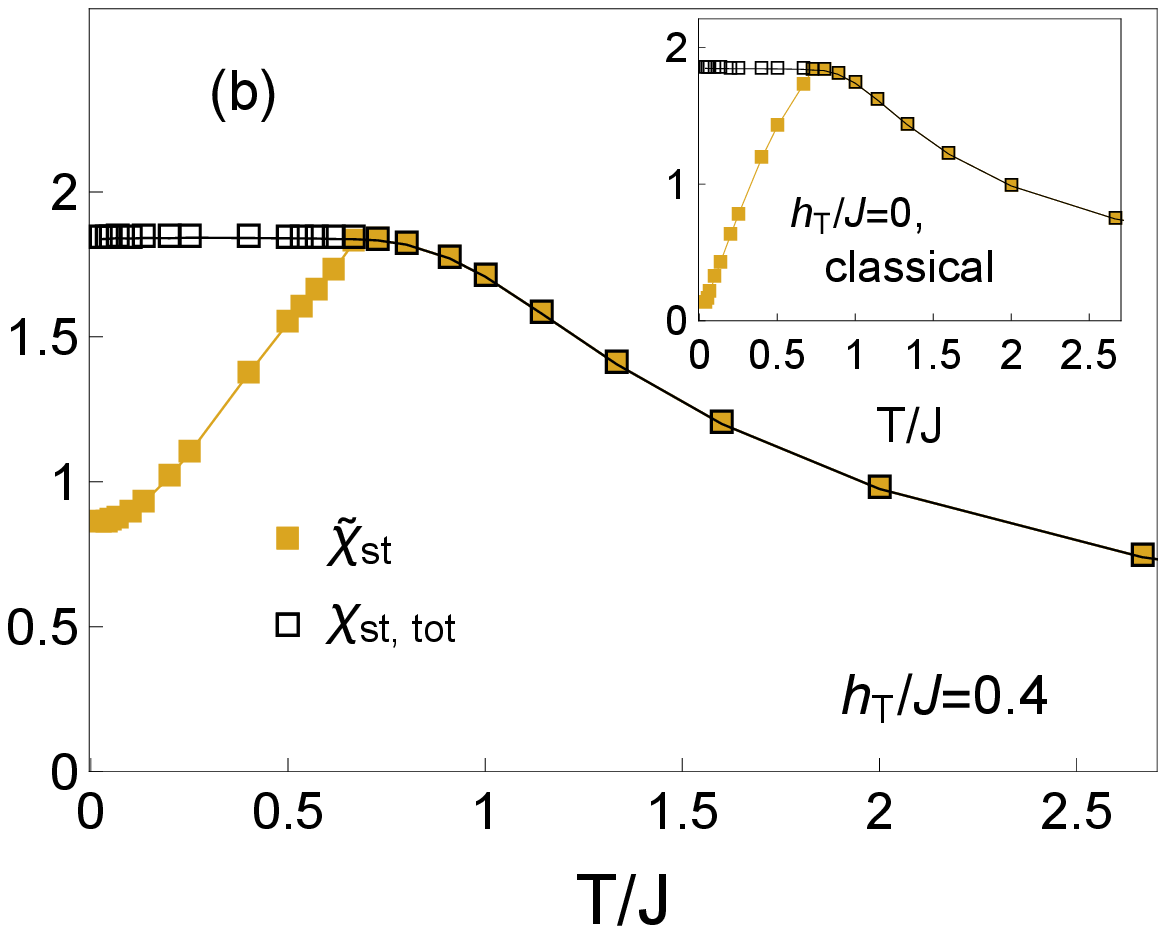}
\caption{Static susceptibilities across the glass transition.
Transverse field (a) and temperature (b) dependence of the susceptibility of a single replica $\tilde\chi_{\rm st}$, and of  the total local susceptibility $ \chi_{\rm st, tot}$. Both susceptibilities coincide in the paramagnetic phase. Single replica susceptibility $\tilde\chi_{\rm st}$ peaks at the transition, and gets strongly suppressed deep in the spin glass phase, while the total $\chi_{\rm st, tot}$ remains smooth and develops a plateau. Qualitatively similar results for the classical limit $h_{\rm T}=0$ are shown in the bottom inset.
}
\label{fig-static-susceptibility}
\end{figure}

While the distributions $P(1,y)$ and $P(Q)$ reveal essential features of the spin glass phase, they are hard to extract in a real physical system. In this section we turn to experimentally  accessible quantities, the static and dynamical spin susceptibilities. We will further explore the experimental relevance of the results presented here in Sec.~\ref{sec:experiment}.

We first explore the static local spin susceptibility of the lattice model ~\eqref{eq:hamiltonian}, expressed as
\begin{equation}\label{eq:chi_phys}
   \chi_{\rm st, tot}= \dfrac{1}{\beta}\dfrac{d^2}{dh_i^2}\log Z,
\end{equation}
with $Z$ denoting the partition function. The subscript `tot' stands for total, for a reason that will become apparent shortly.  By using the replica formula ~\eqref{eq:logZ}, we can rewrite $\chi_{\rm st, tot}$ in terms of the replicated action as
\begin{equation*}
   \chi_{\rm st, tot}=\sum_{b=1}^n\int_\tau \langle\sigma^z_{b\,\tau}\sigma^z_{a\,0}\rangle_{S_{\rm rep}}=\int_\tau\chi(\tau)+\beta \sum_{b:b\neq a} Q_{ab},
\end{equation*}
by using Eq.~\eqref{eq:selfconsistent}.
Performing the replica limit $n\to 0$ results in
\begin{equation*}
   \chi_{\rm st, tot}=\int_\tau\tilde\chi(\tau)+\beta\int dx\left[Q(1)-Q(x)\right].
\end{equation*}
Therefore, in the replica symmetric phase, the total local susceptibility $\chi_{\rm st, tot}$ coincides with the static component of the connected spin correlator within a single replica, $\tilde\chi(\tau)$. In the presence of replica symmetry breaking, however, replica offdiagonal correlators provide correction terms to the total susceptibility. The real space interpretation of this result is that  $\tilde\chi(\tau)$ is the susceptibility of a single metastable state, with further correction terms coming into play as the system has time explore a larger portion of available states and converges to true equilibrium. Susceptibilities measured in a spin glass phase therefore depend sensitively on the experimental details, a phenomenon well known from the difference between field cooled and zero field cooled susceptibilities in glasses.

We display both the susceptibility of a single replica, $\tilde\chi_{\rm st}=\int_\tau\tilde\chi(\tau)$, and the total susceptibility $\chi_{\rm st, tot}$ in Fig.~\ref{fig-static-susceptibility}, as a function of the transverse field $h_{T} $ (a), and of temperature $T$ (b). 
While $\tilde\chi_{\rm st}$ shows a peak at the transition, and becomes strongly suppressed at low $h_T$ and $T$, the  total susceptibility remains smooths across the phase boundary.  Instead, $\chi_{\rm st, tot}$ develops a plateau in the spin glass phase, almost completely insensitive to both $h_T$ and $T$. 
For comparison, the insert of Fig.~\ref{fig-static-susceptibility}b also displays the susceptibilities of the classical Sherrington-Kirkpatrick model, $h_T=0$, showing a behavior very similar to the quantum case, but with a complete suppression of single replica susceptibility, $\tilde\chi_{\rm st}\to 0$ as $T\to 0$.

Now we turn to the discussion of the dynamical properties. Similarly to the static susceptibility, the dynamical local susceptibility can be expressed in the replica formalism.  Since the replica offdiagonal correlations are static, they do not contribute to the $\omega\neq 0$ components of the susceptibility, and the total susceptibility coincides with the single replica susceptibility $\tilde\chi(\omega)$. Here $\tilde\chi(\omega)$ can be calculated numerically by Fourier transforming $\tilde\chi(\tau)$ to the Matsubara frequencies $\Omega_n=2\pi n/\beta$,
\begin{equation*}
    \tilde\chi(\Omega_n)=\int_\tau e^{i\Omega_n\tau}\tilde\chi(\tau),
\end{equation*}
and by performing the analytical continuation to real frequencies, $\Omega_n\rightarrow\omega+i\eta$ with $\eta>0$ infinitesimal, using Pad\'e approximants.

\begin{figure}[t]
\centering
\includegraphics[width=0.8\columnwidth]{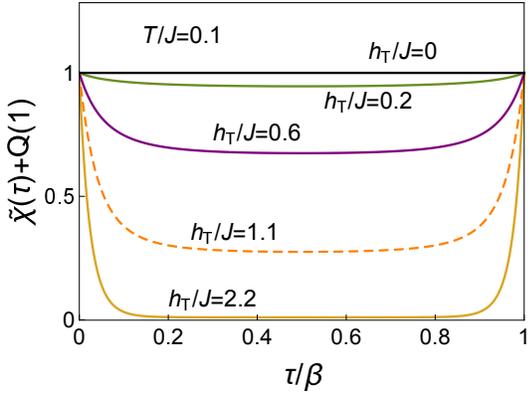}
\caption{
Imaginary-time dependence of the susceptibility $\chi(\tau) = \tilde{\chi}(\tau)$+$Q(1)$, shown for different transverse fields $h_T$.
The classical limit, $h_{\rm T}=0$,  yields a plateau $\chi(\tau)\equiv 1$, with the quantum fluctuations at $h_T\neq 0$ introducing non-trivial imaginary time dependence. The result at the spin glass phase boundary is shown by an orange dashed line.
We used $h_z/J$=0.2, $T/J$=0.1.
}
\label{fig-susceptibility_tau}
\end{figure}

Figure~\ref{fig-susceptibility_tau} shows the imaginary-time dependence of the susceptibility $\chi(\tau)=\tilde{\chi}(\tau)+Q(1)$ for different values of the transverse field $h_{T}$. Here we shifted $\tilde{\chi}(\tau)$ by the constant $Q(1)$, because $\chi(\tau)$  is always normalized as $\chi(0)=\chi(\beta)=1$, allowing a more clear comparison between the results for different fields $h_{T}$. In the classical limit $h_{T}=0$, $\chi(\tau)\equiv 1$, therefore, the deviations from this plateau for $h_T\neq 0$ show the effect of quantum fluctuations. 

In the paramagnetic phase, i.e., for large values of  $h_{T}$ we find that  $\chi(\tau)$ decays exponentially at short times, $\chi(\tau) \sim {\rm exp}(-\tau/\tau_{0})$ with  $\tau_{0} \approx 1/2h_{T}$. Such an exponential  low-$\tau$ behavior is consistent with  the presence of a gap $\Delta\sim 1/\tau_0$ at large frequencies  in the excitation spectrum.
Similar behaviour was presented in Ref.~\onlinecite{rozenberg_grempel1998}.
As we decrease $h_T$ and enter the spin glass phase, $\chi(\tau)$ develops an extended plateau behaviour at large $\tau$ values, appearing after an initial drop in the low-$\tau$ range.

\begin{figure}[t]
\centering
\includegraphics[width=0.9\columnwidth]{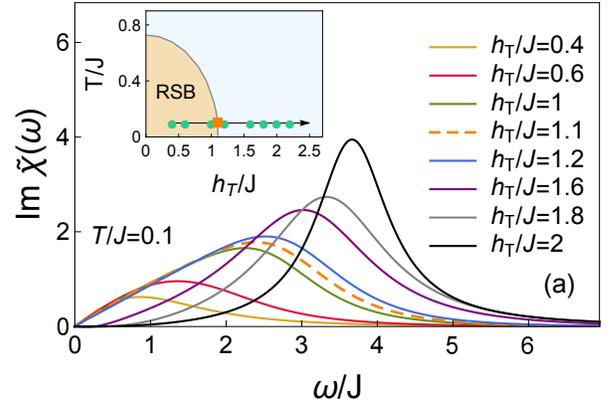}\\ \vspace*{0.5cm}
\includegraphics[width=0.9\columnwidth]{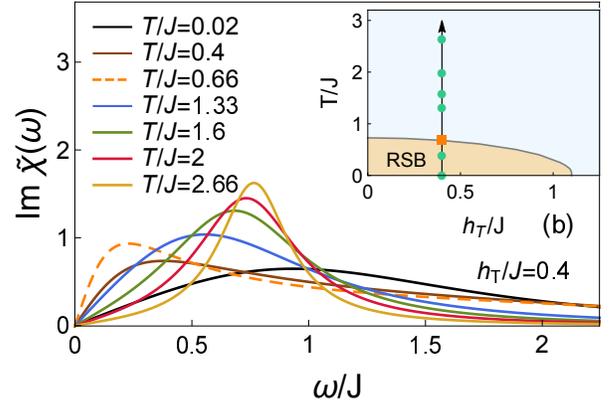}
\caption{Imaginary part of the dynamical susceptibility $\tilde{\chi}(\omega)$ for (a) increasing transverse field $h_T$ and (b) increasing temperature $T$,  moving from the spin glass phase to the paramagnetic phase as shown in the insets. 
Curves at the phase boundary are indicated by orange dashed lines.
We used (a) $h_{z}/J$=0.2, $T/J$=0.1 and (b) $h_{z}/J$=0.2, $h_{T}/J$=0.4. }
\label{fig_Dynamics_RSB}
\end{figure}

The imaginary part of the dynamical susceptibility, Im$\,\tilde\chi(\omega)$, is shown in Fig.~\ref{fig_Dynamics_RSB}, for different pairs of transverse field $h_T$ and temperature $T$ along two distinct  lines crossing to the phase boundary of the spin glass phase.

In Fig.~\ref{fig_Dynamics_RSB}a, $T$ is fixed and the glass transition is crossed by increasing $h_T$, as shown in the inset. Within the spin glass  phase, Im$\,\tilde\chi(\omega)$ increases linearly at small $\omega$ indicating the presence of low energy excitations, in accordance with the plateau found in $\tilde\chi (\tau)$ in Fig.~\ref{fig-susceptibility_tau}. This behavior persists to the critical point (dashed orange line) and into the paramagnetic phase close to phase transition, but   Im$\,\tilde\chi(\omega)$ at small $\omega$ gets   depleted upon increasing $h_T$ further. Deeper in the paramagnetic phase, Im$\,\tilde\chi(\omega)$ reflects the presence of a gap, $\Delta$, also responsible for the exponential short time decay in $\tilde\chi(\tau)$. At large $h_T$, $\Delta\sim 2h_T$ with good precision. 

These results are in good agreement with a $T=0$ exact diagonalization study of the infinite-range random Ising model in a transverse field\cite{rozenberg_arrachea2001}  where direct averages over the disorder were performed, not needing the replica method. 
These simulations also show a substantial transfer of the spectral weight to low frequencies in ${\rm Im}\,\tilde{\chi}(\omega)$ in the spin glass phase as one comes from the paramagnetic phase, i.e., with decreasing the transverse field.
Moreover, the linear $\omega$ dependence of ${\rm Im}\,\tilde{\chi}(\omega)$ in the low-$\omega$ range in the glassy phase is also consistent with our results. 

A different cut  with increasing $T$ and fixed $h_T$, is displayed in Fig.~\ref{fig_Dynamics_RSB}b. The points in parameter space corresponding to the curves of the main figure are again indicated in the inset. Similarly to what we found for increasing quantum fluctuations, Im$\,\tilde\chi(\omega)$ points to the presence of low energy excitations in the spin glass phase, gradually depleted in the paramagnetic phase after crossing the phase boundary (dashed orange line).

\section{Relevance for spin glass experiments}~\label{sec:experiment}

In this section, we comment on the experimental relevance of our mean field quantum spin glass results. The classical Sherrington-Kirkpatrick model has been suggested to give a good qualitative description for  the compound LiHo$_{x}$Y$_{1-x}$F$_4$, derived from the LiHoF$_4$ dipolar-coupled Ising ferromagnet by site dilution with the non-magnetic $Y^{3+}$ ions.
The emerging positional disorder and the frustration of the dipolar interactions drive the system to a classical spin glass phase at low temperatures around doping $x \sim 0.2$ ~\cite{torres2008quantum}.

Applying a transverse magnetic field perpendicular to the easy axis in LiHo$_{x}$Y$_{1-x}$F$_4$ yields a potential realization of a quantum spin glass, qualitatively described by Eq.~(\ref{eq:hamiltonian}). The transverse magnetic field splits the doubly degenerate crystal field ground state of the Ho$^{3+}$ ion with spin-up and spin-down states, by coupling the ground state and the first excited crystal field level, and thereby mixing the classical spin-up and spin-down states. The dynamical properties of the LiHo$_{x}$Y$_{1-x}$F$_4$ compound under such a transverse field have been investigated in a series of a.c. susceptibility measurements~\cite{wu1991classical,Wu_Aeppli1993, Brooke_Aeppli_Sience1999,ghosh2002coherent,torres2008quantum,quilliam2008evidence,quilliam2012exp}, with the strength of quantum fluctuations controlled by the transverse field.

\begin{figure}[t]
\centering
\includegraphics[totalheight=6cm,angle=0]{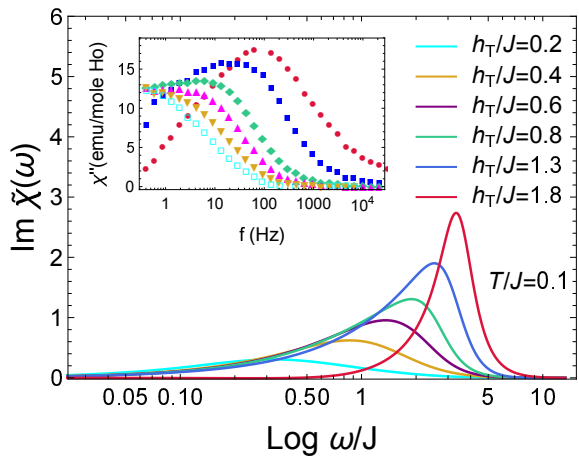}
\caption{Qualitative comparison between mean field quantum spin glass results, and a.c. susceptibility measurements on the LiHo$_{x}$Y$_{1-x}$F$_4$ compound~\cite{torres2008quantum}. Main panel: Imaginary part of the dynamical susceptibility in the quantum Sherrington-Kirkpatrick model, Im$\,\tilde\chi(\omega)$, shown  for several values of transverse field $h_T$ on both sides of the spin glass transition  at $h_{\rm T}^{\rm cr}/J=1.1$. The $h_{T}$ values are chosen as $h_{T}/J=0.2, 0.4, 0.6, 0.8, 1.3, 1.8$ from left to right.
Inset: Results of a.c. susceptibility measurements from Ref.~\onlinecite{torres2008quantum} 
with transverse fields $H_{\rm t}({\rm kOe})=8,6,4,3,2,1$ from right to left.
}
\label{fig_dynamics_comparison_to_experiment}
\end{figure}

In particular, in Ref.~\onlinecite{torres2008quantum}, the frequency dependence of the imaginary part of the dynamical susceptibility, $\chi''(f)$, was measured in LiHo$_{x}$Y$_{1-x}$F$_4$ at $x=0.198$, close to the glass transition, for several values of the transverse field. These measurements pointed towards the formation of a plateau in $\chi''(f)$
in the  spin glass phase at low frequencies $f$, depending only very weakly on the value of the transverse field.
A qualitatively similar behavior for the imaginary part of the dynamical susceptibility, Im$\,\tilde\chi(\omega)$, is obtained in our mean field quantum spin glass calculations shown in Fig.~\ref{fig_Dynamics_RSB}a.
Namely, we find that the low-frequency behavior of Im$\,\tilde\chi(\omega)$ is largely independent of $h_T$ in the spin glass phase, yielding curves that overlap in the low-$\omega$ range, in accordance with experimental results.
This is further illustrated in Fig. ~\ref{fig_dynamics_comparison_to_experiment}, qualitatively comparing our numerical results to the experimental observations of Ref.~\onlinecite{torres2008quantum}.

In contrast to the plateau behavior, remaining robust against variations of the transverse field, $\chi''(f)$ changed non-monotonously in the experiments  as a function of temperature upon  crossing the phase transition in the low-frequency range\cite{torres2008quantum}. 
Similarly to this experimental observation, we also find non-monotonic temperature dependence in the low-frequency limit of Im$\,\tilde\chi(\omega)$, for small values of the transverse field. This is shown in Fig.~\ref{fig_Dynamics_RSB}b above, again in good qualitative agreement with the a.c. susceptibility measurement results reported in Ref.~\onlinecite{torres2008quantum}.

These comparisons demonstrate that the exact solution of simplified mean field quantum spin glass models can already give a lot of insight into the behavior of real materials, and shed light to the qualitative properties of these extremely complex systems.

\section{Outlook to electron glasses}\label{sec:electron}

The theoretical framework presented in this paper, including the quantum Monte Carlo approach, can be extended for fermionic systems, allowing us to obtain the exact solution of mean field electron glass models. 
To demonstrate this, we consider a paradigmatic mean field Coulomb glass model, the disordered $t-V$ model, given by the Hamiltonian
\begin{equation}
\hat{{\cal H}} = -\frac{t}{\sqrt{z}} \sum_{\langle i,j \rangle} (\hat{c}_{i}^{\dag}\hat{c}_{j} + \text{H.c}) + \frac{V}{\sqrt{z}} \delta \hat{n}_{i}  \delta \hat{n}_{j} + \sum_{i} \varepsilon_{i}  \delta \hat{n}_{i}.
\label{eq:CoulombGlassHamiltonian}
\end{equation}
This Hamiltonian describes spinless electrons moving with nearest-neighbor hopping on a Bethe lattice with coordination $z \rightarrow \infty$, experiencing on-site disorder $\varepsilon_{i}$, and interacting with each other through nearest-neighbor repulsive interaction $V_{ij}=V/\sqrt{z}$, mimicking the long-ranged Coulomb interaction. The levels $\varepsilon_{i}$ are drawn from Gaussian distribution $P(\varepsilon) \sim {\rm e}^{-\varepsilon^2/(2W^2)}$, and $\delta \hat{n}_{i}$ denotes deviation from half-filling as $\delta \hat{n}_{i}=\hat{c}_{i}^{\dag}\hat{c}_{i}-1/2$. 

In a previous work~\cite{CoulombGlassPaper2022}, we have already studied the glassy phase of this model by applying  iterative perturbation theory. We also performed exact CTQMC simulations, but only within the replica symmetric Fermi liquid phase.
Here we present the extension of the $h_{T}$-expansion CTQMC algorithm, such that it captures  the exact solution of the mean-field disordered $t-V$ model in both regions, including the electron glass phase with full replica symmetry breaking.
The Monte Carlo algorithm as well as more details on the model are presented in Appendix~\ref{app-CoulombGlass}.

\begin{figure}[t]
\centering
\includegraphics[width=0.9\columnwidth]{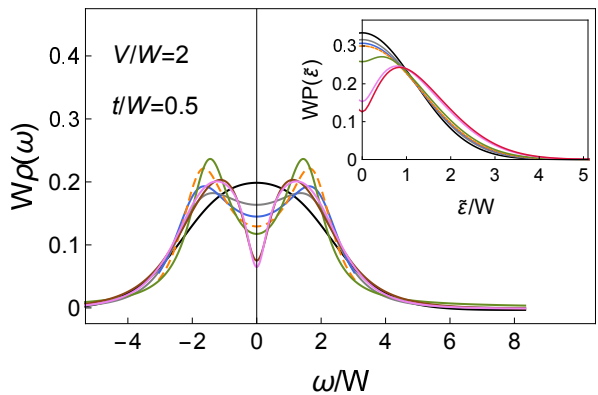}
\caption{Electron glass transition in the disordered $t-V$ mean-field Coulomb glass model. Main panel: Temperature dependence of local density of states across the transition. A correlation hole formed at the Fermi energy in the liquid phase smoothly develops into a pseudogap in the glassy phase. Phase boundary shown by dashed orange line.
Inset: Corresponding distributions of Hartree energies, $P(\tilde{\varepsilon})$, displaying the formation of a pseudogap in the glassy phase.
Temperature chosen as $T=0.5, 0.3, 0.2, 0.156, 0.133, 0.067, 0.05$.
}
\label{fig-CoulombGlass-P}
\end{figure}

We show the evolution of the disorder averaged local density of states $\rho(\omega)$ with decreasing temperature, upon crossing the glass transition, in Fig.~\ref{fig-CoulombGlass-P} (main panel). A correlation hole at the Fermi energy $\omega=0$ starts to appear already in the Fermi liquid phase. After crossing the phase boundary (shown by orange dashed line), this correlation hole develops smoothly into a universal Efros-Shklovskii pseudo-gap deep in the glassy phase. We also show the full distribution of Hartree energies $\tilde{\varepsilon}$, incorporating the bare disorder $\varepsilon_i$ and the renormalization by the interactions. This distribution $P(\tilde{\varepsilon})$ (see inset) is the electron glass analogue of the local magnetic field distribution $P(x=1,y)$ of the Sherrington-Kirkpatrick model, displaying a Gaussian shape in the liquid phase, with a pseudogap starting to open gradually after entering the glassy phase.

\section{Discussion}\label{sec:conclusions}

In this work, we presented the first complete, numerically exact solution of the transverse field Sherrington-Kirkpatrick model, a paradigmatic mean field quantum spin glass model. We combined a continuous-time quantum Monte Carlo method with a replica calculation allowing for full replica symmetry breaking, and gained access to the full phase diagram, as well as to the properties of the glassy phase. We studied in detail several order parameters for the quantum glass transition. Firstly, we examined the distribution of the local effective magnetic field generated by the frustrated Ising interactions, and observed the formation of a pseudo-gap structure in the glassy phase. Secondly, we evaluated and discussed the overlap function $Q_{ab}$, characterizing the overlap in the magnetization patterns of replicas $a$ and $b$ subject to the same disorder. We found that the difference between the maximal and minimal possible overlaps serves as an Edwards-Anderson type order parameter for the transition, and vanishes linearly with the transverse field $h_T$ upon approaching the phase boundary from the glassy phase. We also extracted the full distribution of these overlaps, taking the form of a single Dirac-delta in paramagnetic phase, but broadening to a non-trivial continuous structure upon crossing the glass transition, carrying crucial information on the roughening of the free energy landscape in the spin glass phase.

Turning to experimentally more accessible quantities, we discussed the behavior of the static and dynamical local spin susceptibilities. We found that the static susceptibility develops a flat plateau in the glassy phase, remarkably stable against increasing thermal or quantum fluctuations. The dynamical susceptibility, on the hand, reflects the presence of low energy excitations in the glassy phase, and changes its shape upon crossing the phase boundary. We qualitatively compared these results to a.c. susceptibility measurements performed on the quantum spin glass candidate compound LiHo$_x$Y$_{1-x}$F$_4$, and found a good agreement. These results highlight the relevance of simplified mean field models for understanding the complicated behavior of complex, experimentally accessible materials. 

Besides presenting a detailed study for one of the paradigmatic mean field quantum spin glass models, we demonstrated that the general framework developed in this work is applicable to a wide range of mean-field quantum glass models. To this end, we showed that it can be generalized to capture full replica symmetry breaking in the electron glass phase of a mean field model for Coulomb glasses, the disordered $t-V$ model. These results open up new possibilities to obtain numerically exact solutions for various mean-field quantum glass models, showing different types of orders, such as the quantum Heisenberg spin glass~\cite{HeisenbergGlass_PhysRevLett.85.840}, or the random $t-J$ model relevant for the properties of cuprates~\cite{tJ_PhysRevLett.70.3339,tJ_doi:10.1073/pnas.2206921119}.

Studying the properties of quantum glasses on the mean-field level is an important stepping stone towards understanding real materials. The framework presented in this paper can be combined with a DMFT approach, resulting in a local approximation allowing to investigate realistic, finite dimensional systems. Another exciting open question concerns how the present approach reflects anomalously slow dynamics, and the wide distribution of relaxation time scales. This complex dynamics has potential applications in designing quantum neural networks realizing associative memory, with imminent relevance for ongoing cavity QED experiments~\cite{QED_PhysRevX.11.021048}. Both of these challenging tasks are the subject of future research.

\acknowledgements

We thank C\u{a}t\u{a}lin Pa\c{s}cu Moca for insightful discussions. I.L. acknowledges  support from the Gordon and Betty Moore Foundation through Grant GBMF8690 to UCSB and from the National Science Foundation under Grant No. NSF PHY-1748958. 
This work has been supported by the National Research Development and Innovation Office (NKFIH) through Grant No. K142652.

\appendix

\section{Flow equations in the presence of full replica symmetry breaking}\label{app:flow}

\subsection{Flow equation for the free energy density}

Here we consider the scale dependent free energy density introduced in Sec.~\ref{subsec:rsb}. We first sketch the derivation of the recurrence relation expressing $\phi_{m+\Delta m}(y)$ in terms of $\phi_m(y)$, then we perform the replica limit $n\to 0$, yielding the flow equation ~\eqref{eq:flow_phi}.

A  Parisi block at scale $m+\Delta m$ contains $(m+\Delta m)/m$ blocks of size $m$, which are decoupled under the action $S_{m+\Delta m}(y)$, due to the elimination of the coupling  $Q_{m+\Delta m}$ with a Hubbard-Stratonovich transformation. Therefore, one can  express $S_{m+\Delta m}(y)$ in terms of $1+\Delta m/m$ independent copies of $S_{m}(y)$ as
\begin{align*}
    &S_{m+\Delta m}(y)= \sum_{k=1}^{1+\Delta m/m}S_{m}^{(k)}(y)\\
    &\quad+\dfrac{J^2}{2}\left(Q_{m+\Delta m}-Q_m\right)\sum_{k=1}^{1+\Delta m/m}\sum_{a, b=1}^m \int_\tau\int_{\tau^\prime}\sigma^z_{a\,k\,\tau}\sigma^z_{b\,k\,\tau^\prime},
\end{align*}
with $k$ enumerating the independent $m\times m$ blocks, and $a,b$ distinguishing replicas within a single such block. Therefore, $\phi_{m+\Delta m}(y)$ can be divided into  $1+\Delta m/m$ independent contributions from $m\times m$ blocks,
\begin{widetext}
\begin{align*}
    e^{\beta(m+\Delta m)\phi_{m+\Delta m}(y)}=\left(\int\mathcal{D}\sigma^z\, \exp\left\{-S_m(y)-\frac{J^2}{2}\left(Q_{m+\Delta m}-Q_m\right)\sum_{a, b=1}^m \int_\tau\int_{\tau^\prime}\sigma^z_{a\,\tau}\sigma^z_{b\,\tau^\prime}\right\}\right)^{1+\Delta m/m}
\end{align*}
Moreover, the replica offdiagonal coupling appearing on the right hand side in addition to $S_{m}(y)$ can be eliminated by a Hubbard-Stratonovich transformation as follows,
\begin{align*}
 \beta\phi_{m+\Delta m}(y)&=\dfrac{1}{m} \log \left(\int \dfrac{d\tilde y}{\sqrt{2\pi J^2 \Delta Q_m}}\int\mathcal{D}\sigma^z \exp\left\{-\dfrac{\tilde y^2}{2 J^2 \Delta Q_m}-S_m(y)-\tilde y\sum_{a=1}^m \int_\tau\sigma^z_{a\tau}\right\} \right)\\
 &=\dfrac{1}{m} \log \left(\int \dfrac{d\tilde y}{\sqrt{2\pi J^2 \Delta Q_m}}\int\mathcal{D}\sigma^z \exp\left\{-\dfrac{\tilde y^2}{2 J^2 \Delta Q_m}-S_m(y+\tilde y)\right\} \right)\\
 &=\dfrac{1}{m} \log \left(\int \dfrac{d\tilde y}{\sqrt{2\pi J^2 \Delta Q_m}}\exp\left\{-\dfrac{\tilde y^2}{2 J^2 \Delta Q_m}+\beta\, m\, \phi_m(y+\tilde y)\right\}\right),
\end{align*}
with $\Delta Q_m\equiv Q_m-Q_{m+\Delta m}$. The replica limit $n\to 0$ can now be performed by promoting $m$ to a continuous variable $x\in [0,1]$, and replacing $m+\Delta m$ by $x-\Delta x$, with the sign change stemming from continuing a positive integer $n\geq 1$ to $n=0$. Expanding $\phi_m(y+\tilde y)$ up to second order in $\tilde y$ results in
\begin{align*}
    \beta\left( \phi(x-\Delta x,y)-\phi(x,y)\right)&=\dfrac{1}{x}\log \left(\int \dfrac{d\tilde y}{\sqrt{2\pi J^2 \Delta Q_x }}\exp\left\{-\dfrac{\tilde y^2}{2 J^2 \Delta Q_x }\right\}\left[1+\tilde y^2\,\dfrac{\beta\, x}{2}\left\{\partial_y^2\phi(x,y)+\beta\, x\,(\partial_y\phi(x,y))^2\right\}\right]\right)\\
    &=\dfrac{J^2}{2}\Delta Q_x\, \beta \left\{\partial_y^2\phi(x,y)+\beta\, x\,(\partial_y\phi(x,y))^2\right\}.
\end{align*}
\end{widetext}
Expanding the left hand side to first order in $\Delta x$, and using $\Delta Q_x / \Delta x\rightarrow dQ/dx$ yields the flow equation ~\eqref{eq:flow_phi}.

\subsection{Flow equation for renormalized field distribution}

In this Appendix we give a brief overview of the derivation of the flow equation ~\eqref{eq:flow_P}. The first step is expressing $P_m(y)$ with  $P_{m+\Delta m}(y)$ through a recurrence relation. We consider a spin operator $\mathcal{O}_{m}$ supported within a block at scale $m$. Since $\mathcal{O}_{m}$ is also contained within all larger Parisi blocks of scale $\widetilde m\geq m$, its expectation value can be expressed as
\begin{align*}
    &\langle \mathcal{O}_{m}\rangle_{S_{\rm rep}}=\int dy\, P_{\widetilde m}(y)\,\langle \mathcal{O}_{m}\rangle_{S_{\widetilde m}(y)}=\\
    &\int dy\, P_{\widetilde m}(y)\,e^{-\beta\,\widetilde m\,\phi_{\widetilde m}(y)}\!\!\int_{\widetilde m\times\widetilde m}\!\!\!\!\mathcal{D}\sigma^z\, \mathcal{O}_{m}(\{\sigma^z\})\, e^{-S_{\widetilde m}(y)},
\end{align*}
with $\int_{\widetilde m\times\widetilde m}\mathcal{D}\sigma^z$ denoting a path integral over a spin block of size $\widetilde m\times\widetilde m$, and with $\widetilde m\geq m$ arbitrary. 

We now compare this expression for the subsequent scales $\widetilde m= m$ and $\widetilde m =m+\Delta m$.  Considering $\widetilde m =m+\Delta m$, we note that $S_{m+\Delta m}(y)$ only couples replicas within smaller blocks of size $m\times m$, due to subtracting $Q_{m+\Delta m}$ from the matrix $Q_{ab}$. Therefore, $\int_{(m+\Delta m)\times\ (m+\Delta m)}\mathcal{D}\sigma^z$ factorizes into $(m + \Delta m)/m$ independent path integrals over these smaller blocks, with the operator $\mathcal{O}_{m}$ supported within one of them. The remaining blocks contribute equally towards the total  partition function of the $(m+\Delta m)\times (m+\Delta m)$ block, $e^{\beta\,(m+\Delta m)\,\phi_{ m+\Delta m}(y)}$, leading to a prefactor $e^{\beta\,((m+\Delta m)-m)\,\phi_{ m+\Delta m}(y)}$, with $-m$ in the exponent accounting for the missing contribution from the block containing $\mathcal{O}_{m}$. These considerations yield
\begin{align*}
    &\langle \mathcal{O}_{m}\rangle_{S_{\rm rep}}=\int dy\, P_{m+\Delta m}(y)\,e^{-\beta\,m\,\phi_{m+\Delta m}(y)}\times\\
    &\quad\int_{ m\times m}\!\!\mathcal{D}\sigma^z\, \mathcal{O}_{m}(\{\sigma^z\})\,e^{-S_{m}(y)-J^2 \Delta Q_{m}/2\left(\sum_{a=1}^{m}\int_\tau\sigma^z_{a\,\tau}\right)^2},
\end{align*}
with $\Delta Q_{m}= Q_m- Q_{m+\Delta m}$, where we used the relations from the previous section to express $S_{m+\Delta m}(y)$ within an $m\times m$ Parisi block in terms of $S_{m}(y)$. We can now repeat the steps followed from the calculation of the free energy density, and decouple the replica offdiagonal correction to $S_{m}(y)$ in the exponent with a Hubbard-Stratonovich transformation, by introducing a field $\tilde y$, 
\begin{align*}
    &\langle \mathcal{O}_{m}\rangle_{S_{\rm rep}}=\int dy\, P_{m+\Delta m}(y)\,e^{-\beta\,m\,\phi_{m+\Delta m}(y)}\times\\
    &\qquad\qquad\quad\int \dfrac{d\tilde y}{\sqrt{2\pi J^2 \Delta Q_m}}e^{-\tilde y^2/(2 J^2\Delta Q_{m})}\times\\
    &\qquad\int_{ m\times\ m}\!\!\mathcal{D}\sigma^z\, \mathcal{O}_{m}(\{\sigma^z\})\,e^{-S_{m}(y)-\tilde y\sum_{a=1}^{m} \int_\tau\sigma^z_{a\tau}}\\
    &=\int dy\int \dfrac{d\tilde y}{\sqrt{2\pi J^2 \Delta Q_m}}\,e^{-\tilde y^2/(2 J^2\Delta Q_{m})} P_{m+\Delta m}(y-\tilde y) \\
    &\qquad\qquad\qquad\quad\times\,e^{\beta\,m\,\left[\phi_{m}(y)-\phi_{m+\Delta m}(y-\tilde y)\right]}\langle \mathcal{O}_m\rangle_{S_m(y)}.
\end{align*}
Here the second equality follows from shifting the integration variable, $y\rightarrow y+\tilde y$.

Comparing this relation to the definition of $P_m(y)$ yields a recurrence relation,
\begin{align*}
    P_m(y)=&\int \dfrac{d\tilde y}{\sqrt{2\pi J^2 \Delta Q_m}}\,e^{-\tilde y^2/(2 J^2\Delta Q_{m})}\\
    &\quad\times P_{m+\Delta m}(y-\tilde y)\,e^{\beta\,m\,\left[\phi_{m}(y)-\phi_{m+\Delta m}(y-\tilde y)\right]}.
\end{align*}
Expanding the terms in the second line up to second order in $\tilde y$, and performing the Gaussian integral leads to
\begin{align*}
    &P_m(y)=P_{m+\Delta m}(y)\,e^{\beta\,m\,\left\{\phi_{m}(y)-\phi_{m+\Delta m}(y)\right\}}+\\
    &\dfrac{J^2}{2}\Delta Q_m\!\!\left.\partial_{\tilde{y}}^2\left[P_{m+\Delta m}(y\!-\!\tilde y)\,e^{\beta m\left\{\phi_{m}(y)-\phi_{m+\Delta m}(y-\tilde y)\right\}}\right]\right|_{\tilde y=0}\!.
\end{align*}
One can now proceed by performing the replica limit $m\to x$, $\Delta m\to -\Delta x$ and $\Delta Q_m\to \Delta x\, dQ/dx$, and expanding the right hand side up to order $\Delta x$,
\begin{align*}
    &\partial_x P(x,y)=\\
    &\dfrac{J^2}{2}\dfrac{dQ}{dx}\left[\partial_y^2 P(x,y)-2\beta x\partial_y P(x,y)\partial_y \phi(x,y)\right]+P(x,y)\beta x\\
    &\;\;\times \left[\partial_x\phi(x,y)+\dfrac{J^2}{2}\dfrac{dQ}{dx}\left\{\beta x\left(\partial_y\phi(x,y)\right)^2-\partial_y^2\phi(x,y)\right\} \right].
\end{align*}
The expression in the last line can be simplified by taking into account the flow equation for the free energy density $\phi(x,y)$, yielding the flow equation ~\eqref{eq:flow_P} for $P(x,y)$.

\section{Details of the continuous-time quantum Monte Carlo calculations}\label{app-details-of-CTQMC}

\subsection{Monte Carlo weights of segment configurations}\label{app:ctqmc_weights}

As discussed in the main text, the partition function $Z_y$, Eq.~(\ref{Eq-partition-function-expansion-SpinGlass}), can be expressed as an integral over segment configurations $\boldsymbol{\tau}_{q}$, contributing with weights $w({\boldsymbol{\tau}_{q}})$. According to Eq.~\eqref{eq:weights}, the total weight is expressed as a product of the weights associated with the $y$-field, Eq.~\ref{eq:Szy}, and the interaction term ~\eqref{eq:Schi}, giving rise to the contributions $w_{z}({\boldsymbol{\tau}_{q}},y)$ and $w_{\tilde{\chi}}({\boldsymbol{\tau}_{q}})$, respectively. Below we evaluate these weights factors.

{\sl Evaluation of the $y$-field term.} 
In the following discussions we consider segment configurations with  $\sigma^z_{\tau=0}=-1$. Similar expression hold for $\sigma^z_{\tau=0}=1$.

The trace calculation in Eq.~(\ref{Eq-partition-function-expansion-SpinGlass}) gives rise to the weight factor $w_{z}({\boldsymbol{\tau}_{q}},y)$,
\begin{align}
&w_{z}({\boldsymbol{\tau}_{q}},y)  = {\rm Tr} \left[ {\rm e}^{-S_{z}(y)}\,  \hat{\sigma}^{+}_{\tau^{\prime}_{1}}\,\hat{ \sigma}^{-}_{\tau_{1}} ... \hat{\sigma}^{+}_{\tau^{\prime}_{q}}\, \hat{\sigma}^{-}_{\tau_{q}} \right] = \nonumber \\
&{\rm Tr}\! \left[ {\rm e}^{-y(\beta-\tau_{q})}\, \hat{\sigma}^{-}_{\tau_{q}} \,{\rm e}^{y (\tau_{q}-\tau_{q}^{\prime})} \,
\hat{\sigma}^{+}_{\tau_{q}^{\prime}} \,...\,\hat{\sigma}^{-}_{\tau_{1}} \,
{\rm e}^{y(\tau_{1}-\tau_{1}^{\prime})}\, \hat{\sigma}^{+}_{\tau_{1}^{\prime}} \,{\rm e}^{-y(\tau_{1}^{\prime}-0)}
\right] \nonumber \\
&\qquad\qquad={\rm e}^{-y \left(\ell_{\uparrow}-\ell_{\downarrow}\right)}.
\label{eq-wz_trace_calculation}
\end{align}
Here we used that the only  non-vanishing matrix elements of the spin raising and lowering operators are $ \langle \uparrow |  \hat{\sigma}^{+}_{\tau_{k}^{\prime}} | \downarrow \rangle=\langle \downarrow | \hat{\sigma}^{-}_{\tau_{k}} | \uparrow \rangle=1$ for any $k=1,...,q$. We also defined the total length of segments with spin-up and spin-down states,
$\ell_{\uparrow} = (\tau_{1}-\tau_{1}^\prime) + (\tau_{2}- \tau_{2}^\prime) ... + (\tau_{q}-\tau_{q}^\prime)$ and 
 $\ell_{\downarrow} = (\tau_{1}^\prime-0) +  (\tau_{2}^\prime - \tau_{1}) + ... + (\beta-\tau_{q})$, respectively.

{\sl Evaluation of the interaction term. } To evaluate the interaction term arising from  $S_{\tilde{\chi}}$, Eq.~\eqref{eq:Schi}, it is convenient to introduce an auxiliary function $K(\tau)$, such that $K^{\prime\prime}(\tau) = J^2\,\tilde{\chi}(\tau)$, and the function is periodic in $\beta$ with $K(0)=K(\beta)=0$. The weight $w_{\tilde{\chi}}$ of a given segment configuration can be written as a product over the contributions from all possible pairs of segments $(s_i,s_j)$,
\begin{align}
    w_{\tilde{\chi}}({\boldsymbol{\tau}_{q}})=\prod_{(s_i,s_j)} \exp \left( \frac{J^2}{2} \sigma_i\, \sigma_j \int_{\tau_{i}\in s_i}  \int_{\tau_{j}\in s_j} \tilde{\chi}(\tau_{i}-\tau_{j}) \right),
     \label{Eq-form-of-wtildechi1}
\end{align}
with $\sigma_{i/j}$ denoting the value of $\sigma^z$ on segment $s_{i/j}$. In terms of $K(\tau)$, we obtain,
\begin{eqnarray}
&w_{\tilde{\chi}}({\boldsymbol{\tau}_{q}}) = 
 {\rm exp}\left( \frac{1}{2} \sum_{k_{1},k_{2}} \left[ K(\tau_{k_1} - \tau^{\prime}_{k_2}) + K(\tau^{\prime}_{k_1} - \tau_{k_2}) \right. \right.\nonumber\\
&\quad \left.  \left. - K(\tau_{k_1} - \tau_{k_2}) - K(\tau^{\prime}_{k_1} - \tau^{\prime}_{k_2}) \right]- 2\beta K^{\prime}(0) \right),
\label{Eq-form-of-wtildechi}
\end{eqnarray}
 with $2 K^{\prime}(0)=J^2\int_{\tau} \tilde{\chi}(\tau)$.

\subsection{Monte Carlo procedure}\label{app:Metropolis}

In the CTQMC method, we sample the segment configurations stochastically, by applying a Metropolis algorithm. Here, we decompose the transition probability $W(\boldsymbol{\tau}_{q} \rightarrow \boldsymbol{\tilde\tau}_{\tilde q})$ for moving from a configuration $\boldsymbol{\tau}_{q}$ to a new configuration $\boldsymbol{\tilde\tau}_{\tilde q}$ into a proposal and an acceptance part as $W(\boldsymbol{\tau}_{q}\! \rightarrow\! \boldsymbol{\tilde\tau}_{\tilde q})=W_{\rm prop}(\boldsymbol{\tau}_{q}\! \rightarrow\! \boldsymbol{\tilde\tau}_{\tilde q}) W_{\rm acc}(\boldsymbol{\tau}_{q}\!\! \rightarrow \!\boldsymbol{\tilde\tau}_{\tilde q})$.
The detailed balance condition is satisfied by requiring
\begin{equation}
    W_{\rm acc}(\boldsymbol{\tau}_{q} \rightarrow \boldsymbol{\tilde\tau}_{\tilde q}) = {\rm min}\left(1,\frac{w(\boldsymbol{\tilde\tau}_{\tilde q}) W_{\rm prop}(\boldsymbol{\tilde\tau}_{\tilde q} \rightarrow \boldsymbol{\tau}_{q})}{w(\boldsymbol{\tau}_{q}) W_{\rm prop}(\boldsymbol{\tau}_{q} \rightarrow \boldsymbol{\tilde\tau}_{\tilde q})} \right),
\label{Eq-acceptance-probability}
\end{equation}
with  $w(\boldsymbol{\tau}_{q})$ and $w(\boldsymbol{\tilde\tau}_{\tilde q})$ denoting the weights of the initial and final configurations, respectively.

\begin{figure}
\centering
\includegraphics[totalheight=1.5cm,angle=0]{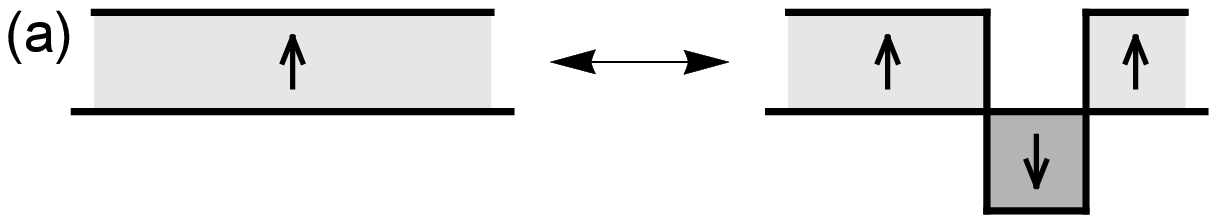}\vspace{0.8cm}
\includegraphics[totalheight=1.5cm,angle=0]{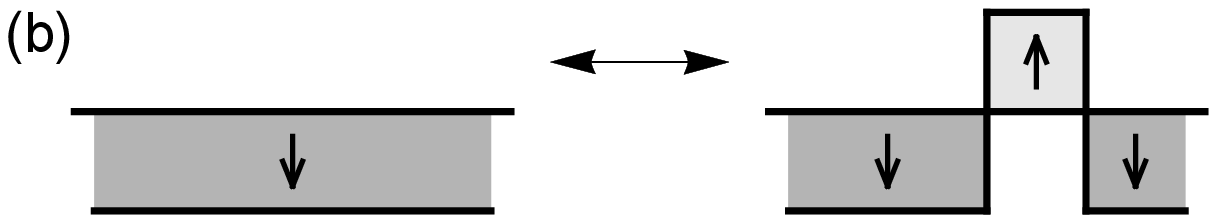}
\caption{Monte Carlo updates for the segment configuration. Addition or removal  of a spin-down segment (a) or of a spin-up segment (b).
}\label{Fig-segments-add-remove-segment}
\end{figure}

We update the segment configuration in the Monte Carlo calculation by either inserting or removing a segment of length $\ell$. As illustrated in Fig.~\ref{Fig-segments-add-remove-segment}, a spin down segment can be added or removed by adding or deleting a pair of neighboring operators $\hat{\sigma}^{+}_{\tau + \ell}\, \hat{\sigma}^{-}_{\tau}$ acting on a spin-up state (Fig.~\ref{Fig-segments-add-remove-segment}a). This operation splits a spin-up ($\sigma^{z}=1$) state into two spin-up states, with an additional spin-down ($\sigma^{z}=-1$) segment in the middle. Similarly, for the insertion or removal of a spin up segment we add or delete  a neighboring pair $\hat{\sigma}^{-}_{\tau^{\prime} + \ell} \hat{\sigma}^{+}_{\tau^{\prime}}$ inside a spin-down segment (Fig.~\ref{Fig-segments-add-remove-segment}b).

A Monte Carlo updating step proceeds as follows.  In the case of the insertion of a spin down segment (adding $\hat{\sigma}^{+}_{\tau + \ell}\, \hat{\sigma}^{-}_{\tau}$), we first select a random imaginary time $\tau$ from the range $\tau \in [0,\beta]$. The left and right endpoints of the segment containing $\tau$ are $\tau_{k}$ and $\tau^{\prime}_{k}$, respectively.
We select a second random imaginary time $\tau^{\prime}$ from the range $\tau^{\prime} \in [0,\ell_{\rm max}]$ with $\ell_{\rm max}=\tau_{k}-\tau$, and the length of the inserted segment is $\ell = \tau^{\prime}-\tau$.
The probability of proposing this operation is $W_{\rm prop}(\boldsymbol{\tau}_{q} \rightarrow \boldsymbol{\tilde\tau}_{ q+1}) = 1/(\beta \ell_{\rm max})$.
The reversed probability  $W_{\rm prop}( \boldsymbol{\tilde\tau}_{ q+1}  \rightarrow \boldsymbol{\tau}_{q})$ corresponds to the removal of a segment. For a removal operation we choose a random segment from the total $q+1$ segments, which gives $W_{\rm prop}( \boldsymbol{\tilde\tau}_{ q+1}  \rightarrow \boldsymbol{\tau}_{q})=1/(q+1)$.
\\
The insertion of a spin up segment (adding $\hat{\sigma}^{-}_{\tau^{\prime} + \ell} \hat{\sigma}^{+}_{\tau^{\prime}}$) goes in a similar manner. Namely, we first select $\tau^{\prime}$ from the range $\tau^{\prime} \in [0,\beta]$, then we select a second imaginary time $\tau$ from the range $\tau \in [0,\ell_{\rm max}]$ with $\ell_{\rm max}=\tau_{k+1}^{\prime}-\tau^{\prime}$ because the segment containing $\tau^{\prime}$ has left endpoint $\tau_{k+1}^{\prime}$ and right endpoint $\tau_{k}$ in this case.
The length of the inserted segment is $\ell = \tau-\tau^{\prime}$.
\\
Using Eqs.~(\ref{Eq-acceptance-probability}) and \eqref{eq:weights},
the acceptance probability of an insertion update $W_{\rm acc}(\boldsymbol{\tau}_{q} \rightarrow \boldsymbol{\tilde\tau}_{ q+1})$ is expressed as
\begin{align*}
&W_{\rm acc}(\boldsymbol{\tau}_{q} \rightarrow \boldsymbol{\tilde\tau}_{ q+1}) = \nonumber \\
& {\rm min}\left( 1,\, \frac{h_{T}^2\, \beta \,\ell_{\rm max}}{\left(q+1\right)} \frac{w_{z} ({\boldsymbol{\tilde\tau}_{q+1}},y)}{w_{z} ({\boldsymbol{\tau}_{q},y})} \frac{w_{\tilde{\chi}} ({\boldsymbol{\tilde\tau}_{q+1}})}{w_{\tilde{\chi}} ({\boldsymbol{\tau}_{q}})} \right)
\end{align*}
with $w_{z} ({\boldsymbol{\tilde\tau}_{q+1}},y)/w_{z} ({\boldsymbol{\tilde\tau}_{q},y})=e^{2 y(\tau^{\prime}-\tau)}$, where
$\tau^{\prime}-\tau = \ell$ for addition of a spin down segment, and $\tau^{\prime}-\tau = -\ell$ for addition of a spin up segment.
\\
With a similar consideration, the acceptance probability for a segment removal update is given by
\begin{align*}
&W_{\rm acc}(\boldsymbol{\tau}_{q} \rightarrow \boldsymbol{\tilde\tau}_{ q-1}) = {\rm min}\left( 1,\, \frac{q\,e^{-2 y(\tau^{\prime}-\tau)}}{h_{T}^2\, \beta \,\ell_{\rm max}} \frac{w_{\tilde{\chi}} ({\boldsymbol{\tilde\tau}_{q-1}})}{w_{\tilde{\chi}} ({\boldsymbol{\tau}_{q}})} \right).
\end{align*}

Both update probabilities are always positive, therefore, the calculations do not suffer from negative sign problem.

\subsection{Operator expectation values}\label{app:observables}

The contribution of a given segment configuration to the average magnetization $\langle \sigma^{z} \rangle_{y}$ is given by
\begin{equation}
    \langle \sigma^{z} \rangle_{y} = \dfrac{1}{\beta}\sum_{s\in{\rm segments}}\sigma_s\,\ell_s=  \dfrac{1}{\beta}(\ell_{\uparrow} - \ell_{\downarrow}),
    \label{eq-sigmaz-ctqmc}
\end{equation}
with $\ell_s$ denoting the length of segment $s$, with $\sigma_s=\pm$ for a spin up / down state.

To evaluate the connected correlator $\tilde{\chi}_{y}(\tau)$, in the CTQMC it is more convenient to consider the correlation function
\begin{equation*}
    \chi_{y}(\tau)=\tilde{\chi}_{y}(\tau)+\langle \sigma^{z} \rangle_{y}^2=\langle T_{\tau} \sigma^{z}_{\tau} \sigma^{z}_{0} \rangle_{S(y)}.
\end{equation*}
The contribution of a segment configuration to $\chi_{y}(\tau)$ can be evaluated by shifting the segment configuration around the circle of circumference $\beta$ by $\tau$. For a segment $s$ corresponding to the arc $\{\tau_i,\tau_j\}$, this yields a shifted segment $s_\tau$ on the arc $\{\tau_i+\tau \mod{\beta},\tau_j+\tau \mod{\beta}\}$. Using this notation, we obtain
\begin{equation}
    \chi_{y}(\tau)=\dfrac{1}{\beta^2}\sum_{s,s^\prime\in{\rm segments}}\sigma_s\sigma_{s^\prime}\,\ell_{s\cap s_{\tau}^{\prime}},
    \label{eq-chi-ctqmc}
\end{equation}
with $\ell_{s\cap s_{\tau}^{\prime}}\geq 0$ denoting the length of the intersection between the segment $s$ and the shifted segment $s_{\tau}^{\prime}$.


\section{Monte Carlo algorithm for the disordered $t-V$ model}\label{app-CoulombGlass}

In this section we present the extension of the $h_{T}$-expansion Monte Carlo algorithm to the case of the disordered $t-V$ model in the mean field limit, $z \rightarrow \infty$.
We start with a short summary of the theoretical background, and then present the Monte Carlo algorithm.
A detailed discussion of the model is given in Ref.~\onlinecite{CoulombGlassPaper2022}.

\subsection{Theoretical background}

Following similar steps in the replica formalism as it is outlined in Section~\ref{subsec:replica}, 
the local replicated effective action in the $z \rightarrow \infty$ mean-field limit of the disordered $t-V$ model given with the Hamiltonian~(\ref{eq:CoulombGlassHamiltonian}) is obtained as
\begin{widetext}
\begin{align}\label{eq:Srep_f}
S_{\rm rep}=\int_0^\beta d\tau\int_0^\beta d\tau^\prime&\Bigg\lbrace\sum_{a=0}^n\left(\overline{c}_{\,\tau}^{\,a}\left[\delta(\tau-\tau^\prime)\partial_{\tau^\prime}-t^2 G(\tau-\tau^\prime)\right] {c}_{\,\tau'}^{\,a}-\dfrac{V^2}{2}\chi(\tau-\tau^\prime)
\delta n_{\,\tau}^{\,a} \; \delta n_{\,\tau'}^{\,a} \right)\nonumber\\
&\qquad-\dfrac{1}{2}\sum_{a\ne b }^n\ V^2 Q_{ab} \delta n_{\,\tau}^{\,a} \delta n_{\,\tau'}^{\,b} 
 -  \dfrac{1}{2} \sum_{a,b=0}^n  W^2    \delta n_{\,\tau}^{\,a} \delta n_{\,\tau'}^{\,b} 
\Bigg\rbrace,
\end{align} 
supplemented by the self-consistency conditions\begin{align}\label{eq:consistency2}
G(\tau-\tau^\prime)  =\left\langle c^{\,a}_\tau\, \overline c ^{\,a}_{\tau^\prime}\right\rangle_{_{S_\text{rep}}},\quad\quad
\chi(\tau-\tau^\prime)&=\left\langle\delta n^a_\tau\,  \delta n^a_{\tau^\prime}  
\right\rangle_{S_{\rm rep}},\quad\quad
Q_{a\ne b}=
\large\langle\delta n^a_\tau  \,  \delta n^b_{\tau^\prime}  \large\rangle_{S_\text{rep}}\,.
\end{align}
\end{widetext}
Here, the glass order parameter $Q_{ab}$ expresses density fluctuation correlations between different replicas.

As it is in the case of the SK model, in the replica symmetric solution the $Q_{a\ne b}=Q_{RS}$ Ansatz is assumed, which solution describes a disordered Fermi liquid phase.   
The local effective action $S_{\tilde{\varepsilon}}$ is obtained from $S_{\rm rep}$ by decoupling different replicas by the Hubbard-Stratonovitch field $\tilde{\varepsilon}$, giving
\begin{eqnarray}
S_{\tilde{\varepsilon}} &=& \int_{\tau} \int_{\tau^{\prime}} \left\{ \overline{c}_{\tau} \left[ \delta_{\tau, \tau^{\prime}} \left[ \partial_{\tau^{\prime}} + \tilde{\varepsilon} \right] - t^2 G(\tau-\tau^{\prime})\right] c_{\tau^{\prime}} \right. \nonumber \\
&-& \left. \frac{V^2}{2} \left( \chi(\tau - \tau^{\prime}) - Q_{RS} \right) \delta n_{\tau} \delta n_{\tau^{\prime}} -\frac{\beta \tilde{\varepsilon}}{2}
\right\}.
\label{Eq-effective-action-CoulombGlass}
\end{eqnarray}
The Hubbard-Stratonovic fields $\tilde{\varepsilon}$ have a  Gaussian distribution as
$P_{RS}(\tilde{\varepsilon})\sim  \exp\big(-\tilde{\varepsilon}^2/(W^2 + V^2Q_\text{RS})/2\big)$.
The self-consistency conditions in Eq.~(\ref{eq:consistency2}) become
 \begin{equation}
\left\{\begin{array}{c}
       G(\tau) \\
{\chi}(\tau)
        \end{array}\right\}
        = \int d \tilde{\varepsilon} \,  P_{RS}(\tilde{\varepsilon}) \,
      \left\{\begin{array}{c}  
        {G}_{\tilde{\varepsilon}}(\tau\,)\\
    { \chi}_{\tilde{\varepsilon}}(\tau\,)
        \end{array}\right\}\,,
        \label{eq:average}
\end{equation}
and $Q_{RS}$ is also determined self-consistently by  
\begin{equation*}
Q_{RS} = \overline{ \langle\delta n\rangle^2} =  \int \text{d}\tilde{\varepsilon}\, P_{RS}(\tilde{\varepsilon}) \; {\langle \delta n\rangle_{\tilde{\varepsilon}}}^{2}\;. 
 \end{equation*}
The quantitites ${G}_{\tilde{\varepsilon}}(\tau\,)$,  ${ \chi}_{\tilde{\varepsilon}}(\tau\,)$, and  $ \langle\delta n\rangle_{\tilde{\varepsilon}}$ are computed with the effective local action, 
Eq.~\ref{Eq-effective-action-CoulombGlass}. 

In the case of  full replica symmetry breaking, the local effective action (\ref{Eq-effective-action-CoulombGlass}) still holds with the self-consistency conditions~(\ref{eq:average}), only the substitution $Q_{RS} \rightarrow Q_{aa}$ should be taken.
However, the distribution $ P_{RS}(\tilde{\varepsilon})$ of the local energy levels will be deformed from Gaussian form to a more complicated, non-Gaussian structure that should be determined self-consistently.
This is achieved by solving the flow equations which have the form
\begin{align*}
& \frac{\partial \phi_{x,\tilde{\varepsilon}}}{\partial x}
=-\dfrac{V^2}{2}\frac{dQ}{dx}\left\lbrace  \frac{\partial^{2} \phi_{x,\tilde{\varepsilon}}}{\partial \tilde{\varepsilon}^2}+\beta x\left(  \frac{\partial \phi_{x,\tilde{\varepsilon}}}{\partial \tilde{\varepsilon}}  \right)^2\right\rbrace, 
\\
& \frac{\partial P_{x,\tilde{\varepsilon}}}{\partial x} =\dfrac{V^2}{2}\frac{dQ}{dx}\left\lbrace \frac{\partial^{2} P_{x,\tilde{\varepsilon}}}{\partial \tilde{\varepsilon}^2}-2\beta x \frac{\partial}{\partial \tilde{\varepsilon}} \left(  P_{x,\tilde{\varepsilon}} \frac{\partial \phi_{x,\tilde{\varepsilon}}}{\partial \tilde{\varepsilon}} \right)
\right\rbrace,
\end{align*}
where we used the notations $P_{x,\tilde{\varepsilon}} \equiv P(x,\tilde{\varepsilon})$ and $\phi_{x,\tilde{\varepsilon}} \equiv \phi(x,\tilde{\varepsilon})$. The flow equations are subject to the  boundary conditions that $\phi_{1,\tilde{\varepsilon}}$ is the free energy of the replica diagonal action, and $P_{0,\tilde{\varepsilon}}$ takes a Gaussian form similar to $P_{RS}$, with the substitution $Q_{RS}\rightarrow Q_0$.

For more details, please visit Ref.~\onlinecite{CoulombGlassPaper2022}.

\subsection{Monte Carlo algorithm}

\begin{figure}[t]
\centering
\includegraphics[width=0.8\columnwidth]{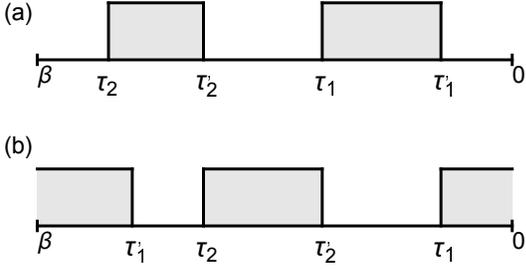}
\caption{CTQMC configurations with expansion order $q=2$ for the case of the Coulomb glass problem corresponding to operator sequences $ \hat{c}^{\dag}_{\tau_{1}^{\prime}}\hat{c}_{\tau_{1}} ...  \hat{c}^{\dag}_{\tau_{2}^{\prime}}\hat{c}_{\tau_{2}}$ (a) and $ \hat{c}_{\tau_{1}}\hat{c}^{\dag}_{\tau_{2}^{\prime}} ...  \hat{c}_{\tau_{2}}\hat{c}^{\dag}_{\tau_{1}^{\prime}}$ (b) contributing to the partition function in the hybridization-expansion, visualized in the segment picture.
}\label{Fig-segments-CoulombGlass}
\end{figure}

The computation of $\langle \delta n \rangle_{\tilde{\varepsilon}}$, $G_{\tilde{\varepsilon}}(\tau)$, and the susceptibility $\chi_{\tilde{\varepsilon}}(\tau)$ with the local effective action~(\ref{Eq-effective-action-CoulombGlass})
 can be performed by the CTQMC.
The Monte Carlo algorithm is a version of the $h_{T}$-expansion algorithm presented in Section~\ref{sec:ctqmc} and Appendix~\ref{app-details-of-CTQMC} by treating fermions instead of the Ising spin variables. 

Here, instead of the expansion in $h_{T}$,
we expand the partition function $Z_{\tilde{\varepsilon}} = {\rm Tr}{\rm e}^{-S_{\tilde{\varepsilon}}}$ in terms of the hybridization function $F(\tau - \tau^{\prime})= t^2G(\tau - \tau^{\prime})$.
The expansion reads as
\begin{eqnarray}
Z_{\tilde{\varepsilon}} &=& {\rm Tr}\,{\rm e}^{-S_{\rm F}+S_{1}} \nonumber \\
&=&   \sum_{q} \int_{\tau} \int_{\tau^{\prime}} {\rm det} \hat{F}^{(q)} {\rm Tr} \left[ {\rm e}^{-S_{1}} 
\overline{c}_{\tau_{1}} c_{\tau^{\prime}_{1}} \dots \overline{c}_{\tau_{q}} c_{\tau^{\prime}_{q}} \right] \nonumber \\
\label{Eq-partition-function-expansion-CoulombGlass}
\end{eqnarray}
with the action terms $S_{\rm F}\equiv - \int_{\tau} \int_{\tau^{\prime}}  \overline{c}_{\tau}  t^2 G(\tau-\tau^{\prime})c_{\tau^{\prime}} $ and $S_{1} \equiv \int_{\tau} \overline{c}_{\tau}(\partial_{\tau} + \tilde{\varepsilon}) c_{\tau} - V^2/2\int_{\tau} \int_{\tau^{\prime}} \tilde{\chi}(\tau-\tau^{\prime}) \delta n_{\tau} \delta n_{\tau^{\prime}}$,
where $\tilde{\chi}(\tau-\tau^{\prime}) = \chi(\tau-\tau^{\prime}) -Q(1) $.
The matrix $\hat{F}^{(q)}$ in Eq.~(\ref{Eq-partition-function-expansion-CoulombGlass}) is composed of the hybridization functions as $\hat{F}^{(q)}_{ji} = F(\tau_{i} - \tau^{\prime}_{j}) = t^2 G(\tau_{i} - \tau^{\prime}_{j}) $.
The partition function can be expressed as an integral over configurations,
$Z_{\tilde{\varepsilon}}  =  \int {\cal D}( {\boldsymbol{\tau}_{q}}) w({\boldsymbol{\tau}_{q}})$, where a configuration ${\boldsymbol{\tau}_{q}}$ is a set of imaginary times, ${\boldsymbol{\tau}_{q}} = \{ \tau^{\prime}_{1}, \tau_{1}, ... \tau^{\prime}_{q}, \tau_{q} \}$, as it was in the $h_{T}$-expansion method as well.

Figure~\ref{Fig-segments-CoulombGlass} shows the the segment representation of the configurations which takes a series of segments $\{ \tau^{\prime}_{k}, \tau_{k} \}$ on which the particle number (occupation) is 1, and 0 otherwise.
The weight $w({\boldsymbol{\tau}_{q}})$  is expressed as  $w({\boldsymbol{\tau}_{q}}) = {\rm det} \hat{F}^{(q)} w({\boldsymbol{\tau}_{q}}, \tilde{\varepsilon}) \tilde{w}_{\tilde{\chi}}({\boldsymbol{\tau}_{q}})$, where the weight factors $ w({\boldsymbol{\tau}_{q}}, \tilde{\varepsilon})$ and $\tilde{w}_{\tilde{\chi}}({\boldsymbol{\tau}_{q}})$ come from the level energies $\tilde{\varepsilon}$ and Coulomb interaction, respectively.
The weight factor $\tilde{w}_{\tilde{\chi}}({\boldsymbol{\tau}_{q}})$ has the same form as $w_{\tilde{\chi}}({\boldsymbol{\tau}_{q}})$ given in Eq.~\ref{Eq-form-of-wtildechi1} by substituting the Ising interaction $J$ with the Coulomb interaction $V$.
Derivation of $ w({\boldsymbol{\tau}_{q}}, \tilde{\varepsilon})$ goes in a similar way as it is outlined in Eq.~(\ref{eq-wz_trace_calculation}) for the spin glass problem because of the formal correspondence $\tilde{\varepsilon} \sim y$. Namely we obtain
$w({\boldsymbol{\tau}_{q}}, \tilde{\varepsilon})= {\rm e}^{-\tilde{\varepsilon} \ell}$, where $ \ell=\sum_{i=1}^{q} \ell_{i}$ with  $ \ell_{i} = \tau_{i} - \tau^{\prime}_{i}$ is the total length of the segments.

A given segment configuration ${\boldsymbol{\tau}_{q}}$ contributes to the expectation value $\langle \delta n \rangle_{\tilde{\varepsilon}} = \langle n \rangle_{\tilde{\varepsilon}} - 1/2$ through the occupation number $ \langle n \rangle_{\tilde{\varepsilon}}$ which is evaluated as
\begin{eqnarray}
 \langle n \rangle_{\tilde{\varepsilon}} = \frac{1}{\beta} \sum_{i=1}^{q} \ell_{i}=\frac{1}{\beta}  \ell.
\label{eq-n-CoulombGlass}
\end{eqnarray}
To compute the contribution of a segment configuration to the susceptibility $\chi_{\tilde{\varepsilon}}(\tau) = \langle T_{\tau} \delta n_{\tau} \delta n_{0} \rangle_{\tilde{\varepsilon}}$, it is convenient to express it as
\begin{eqnarray}
\chi_{\tilde{\varepsilon}}(\tau) =  \langle T_{\tau} n_{\tau} n_{0} \rangle_{\tilde{\varepsilon}} - \langle n \rangle_{\tilde{\varepsilon}} +\frac{1}{4}.
\label{eq-chi-CoulombGlass}
\end{eqnarray}
The second term in the right hand side of Eq.~(\ref{eq-chi-CoulombGlass}) is evaluated on a segment configuration as it is given in Eq.~(\ref{eq-n-CoulombGlass}),
while the first term as 
\begin{eqnarray}
\langle T_{\tau} n_{\tau} n_{0} \rangle_{\tilde{\varepsilon}} = \frac{1}{\beta^2} \sum_{s,s^\prime\in{\rm segments}}\ell_{s\cap s_{\tau}^{\prime}}
\end{eqnarray}
with $\ell_{s\cap s_{\tau}^{\prime}}$ denoting the length of the intersection between the segment $s$ and the shifted segment $s_{\tau}^{\prime}$.
Finally, the calculation of the Green's function is performed as
\begin{eqnarray}
G(\tau) = \left\langle \frac{1}{\beta} \sum^{q}_{i,j} \left( \hat{F}^{(q)} \right)^{-1}_{ji} \delta(\tau, \tau_{i} - \tau^{\prime}_{j})  \right\rangle_{\rm MC},
\end{eqnarray}
where $\hat{F}^{(q)}_{ji} = F(\tau_{i} - \tau^{\prime}_{j})$, and
$\delta(\tau, \tau_{i} - \tau^{\prime}_{j}) = \delta(\tau - \tau^{\prime})\,\, {\rm if}\,\,\tau^{\prime}>0$ while  $\delta(\tau, \tau_{i} - \tau^{\prime}_{j}) = - \delta(\tau - \tau^{\prime})\,\, {\rm if}\,\,\tau^{\prime}<0$.


\bibliography{references}

\end{document}